\documentclass[lettersize,journal]{IEEEtran}
\usepackage{algpseudocode}
\usepackage{algorithmicx}
\usepackage[ruled,vlined,linesnumbered]{algorithm2e}
\usepackage{amssymb}
\usepackage{booktabs}
\usepackage{graphicx, cite}
\usepackage{subfigure}
\usepackage{amsmath,bm}
\usepackage{amsthm}
\usepackage{float}
\usepackage{stfloats}
\usepackage{url}
\usepackage{multirow}
\usepackage{xcolor}
\usepackage{CJK}
\usepackage{mathrsfs}
\usepackage{enumitem}
\graphicspath{{figures/}}

\SetCommentSty{mycommfont}

\IEEEoverridecommandlockouts

\newcommand{\journal}[1]{{\color{black}{#1}}}
\newcommand{\revision}[1]{{\color{black}{#1}}}

\begin{document}
\title{\textit{FAIR:} Towards Impartial Resource Allocation for Intelligent Vehicles with Automotive Edge Computing} 

\author{Haoxin Wang,~\IEEEmembership{Member,~IEEE,}
Jiang (Linda) Xie,~\IEEEmembership{Fellow,~IEEE,} and Muhana Magboul Ali Muslam,~\IEEEmembership{Member,~IEEE}\\

\thanks{Haoxin Wang is with Georgia State University, Department of Computer Science, 25 Park Place, Atlanta, GA 30303 (email: haoxinwang@gsu.edu); 

Jiang (Linda) Xie is with The University of North Carolina at Charlotte, Department of Electrical and Computer Engineering, 9201 University City Blvd, Charlotte, NC 28223 (email: linda.xie@uncc.edu);

Muhana Magboul Ali Muslam is with Imam Mohammad Ibn Saud Islamic University, Department of Information Technology, Riyadh 11432, Saudi Arabia (email: mmmuslam@imamu.edu.sa).
}}

\maketitle

\begin{abstract}
The emerging vehicular connected applications, such as cooperative automated driving and intersection collision warning, show great potentials to improve the driving safety, where vehicles can share the data collected by a variety of on-board sensors with surrounding vehicles and roadside infrastructures. Transmitting and processing this huge amount of sensory data introduces new challenges for automotive edge computing with traditional wireless communication networks. In this work, we address the problem of traditional asymmetrical network resource allocation for uplink and downlink connections that can significantly degrade the performance of vehicular connected applications. An end-to-end automotive edge networking system, \textit{FAIR}, is proposed to provide \underline{f}ast, sc\underline{a}lable, and \underline{i}mpa\underline{r}tial connected services for intelligent vehicles with edge computing, which can be applied to any traffic scenes and road topology. The core of \textit{FAIR} is our proposed symmetrical network resource allocation algorithm deployed at edge servers and service adaptation algorithm equipped on intelligent vehicles. Extensive simulations are conducted to validate our proposed \textit{FAIR} by leveraging real-world traffic dataset. Simulation results demonstrate that \textit{FAIR} outperforms existing solutions in a variety of traffic scenes and road topology.
\end{abstract}

\begin{IEEEkeywords}
Connected and Automated Vehicles, Edge Computing, Intelligent Driving
\end{IEEEkeywords}


\section{Introduction}
\label{sc:introduction}
\IEEEPARstart{L}{ess} \journal{than half a decade, the expeditious evolution of wireless communications, Artificial Intelligence (AI), and high performance computing (HPC), has been reinventing the mobility concept and systems. For example, in 2019, Toyota announced a profound transformation from being an automaker to become a mobility company, with an emphasis on \underline{C}onnectivity, \underline{A}utonomous driving, \underline{S}hared mobility, and \underline{E}lectrification of vehicles (CASE) \cite{ToyotaCASE}. Meanwhile, other major automotive original equipment manufacturers (OEMs), including Volkswagen, Audi, BMW, etc., are investing heavily in future mobility solutions to enhance their core competencies and to ingratiate themselves with their customers. In addition, a variety of cutting-edge technologies and applications have been proposed to facilitate CASE and next-generation intelligent vehicles, such as mobility digital twins (MDT) \cite{wang2022mobility}, automotive edge computing \cite{liu2019edge, lu2022comparison, 9827596}, and vehicle-to-everything (V2X) communications with 5G \cite{garcia2021tutorial,9726924, 9645261,9291414}.
}

\journal{The mobility digital twin is an emerging implementation of Digital Twins (DT) in the transportation domain and is one of the killer applications of the CASE. The MDT aims to create a perpetual digital replica of a human driver or a connected and automated vehicle (CAV) in the digital world based on the data acquired in the physical world. It has been attracting tremendous attention from both industry and academia. According to a market research report, the global DT market size was valued at USD $7.48$ billion in 2021 and is projected to grow at a compound annual growth rate of 39.1\% from 2022 to 2030, where the automotive and transportation industry accounted for the largest revenue share of more than 19.0\% of the overall revenue \cite{grand2022digital}. Very recently, a few MDT research have investigated and demonstrated how the MDT can facilitate next-generation CAVs and potential connected services, including cooperative automated driving \cite{wang2020architectural,9497693, 9857660, 9415170}, personalized adaptive cruise control (P-ACC) \cite{wang2022gaussian, 9152161}, driving behavior modelling and prediction \cite{hu2022review, 9511277, 9259200}, and 3D virtual environment for autonomous driving testing \cite{dRisk}. One of the shared attributes among all these MDT-assisted connected services is the requirement of continuous physical sensory data collection, modelling, and data processing. For example, $4$ terabytes of sensory data will be gathered in two hours from a CAV's on-board camera, LiDAR, and radar sensors \cite{Intel, Darwish_2018_Access}. The volume of the data collected by perception sensors on CAVs would be extremely huge even in a scale of ten vehicles, which requires a large amount of computation resource. Therefore, it is imperative to investigate how to efficiently and fast collect and process these sensory data.  

One of the promising solutions to this challenge is automotive edge computing, which brings the computing and storage resources at the location where the data is generated. Instead of processing these huge amount of sensory data on vehicles or sending to a remote cloud server, the data can be transmitted to and processed on an edge server that is in close proximity to vehicles. For example, an edge server can be deployed with a roadside unit (RSU), and vehicles can offload their sensory data directly to the edge server in real time through wireless communication technologies. In addition, the edge server can provide the capability of real-time data processing and data caching for those emerging MDT-assisted connected services.

\textbf{Motivation.} Although compared to processing sensory data locally on a vehicle or transmitting the data to a remote cloud server, edge-assisted solutions can enhance the efficiency and reduce the end-to-end latency (i.e., starting from the data collection to receiving the processing result), it still remains challenges of providing reliable, scalable, and resilient edge computing networks for vehicles with dynamics in data offloadings. Furthermore, the quality of service (QoS) of vehicular connected services in edge computing networks cannot always be guaranteed due to the legacy design of radio resource allocation in existing wireless networks. 

Traditionally, wireless networks allocate more radio resources to downlink, since downlink traffic volume is usually much higher than uplink. For example, downloading an entertainment video from the YouTube server consumes much more bandwidth than uploading a vehicle's GPS information for the navigation purpose. Therefore, downlink acquires higher throughput, peak data rate, and spectral efficiency than uplink. However, MDT-assisted connected services may require a higher throughput than the vehicular downlink traffic. For instance, a CAV with cooperative automated driving needs to continuously offload its on-board camera captured image frames to an edge server for further processing in real time, while the edge server will send back the processing results (downlink traffic) to the CAV. The uplink traffic will be massive and latency-sensitive compared to the downlink traffic, and thus requires much more network resource than the downlink traffic. \textit{Such new traffic distribution creates unique challenges for MDT-assisted connected services in automotive edge computing and requires the radio resource allocation issue in wireless networks to be revisited and redesigned.}
}

\journal{\textbf{Our contributions.} In this paper, we study these research challenges in automotive edge computing networks and design impartial resource allocation algorithms for MDT-assisted connected services. The novel contributions of this work are summarized as follows: 
\begin{enumerate}
    \item This work is, to the best of our knowledge, the first to systematically address the asymmetrical network resource allocation for uplink and downlink connections in automotive edge computing networks.
    \item We implement a physical network testbed to conduct preliminary experimental study and to demonstrate the resource allocation issues in traditional wireless communication networks.
    \item We propose and design an end-to-end automotive edge networking system, \textit{FAIR}, which provides \underline{f}ast, sc\underline{a}lable, and \underline{i}mpa\underline{r}tial connected services for intelligent vehicles with edge computing in any traffic scenes.
    \item We conduct extensive data-driven simulations to validate our proposed method based on the real world traffic data with different road topology and traffic scenes.
\end{enumerate}

The remainder of this paper is organized as follows. Section \ref{sc:motivation} introduces our preliminary study and the observations from the experiments. Section \ref{sc:Proposed Scheme} presents the analytical factors in edge-assisted intelligent vehicle systems and the details of the proposed \textit{FAIR} that consists of a symmetrical resource allocation algorithm and a service adaption algorithm. Section \ref{sc:evaluation} evaluates the performance of \textit{FAIR} through extensive simulations with real-world traffic data. Finally, the conclusion is drawn in Section \ref{sc:Conclusion}.
}

\section{Preliminary Experiments}
\label{sc:motivation}

\begin{figure*}[t]
\centering
\includegraphics[width=0.88\textwidth]{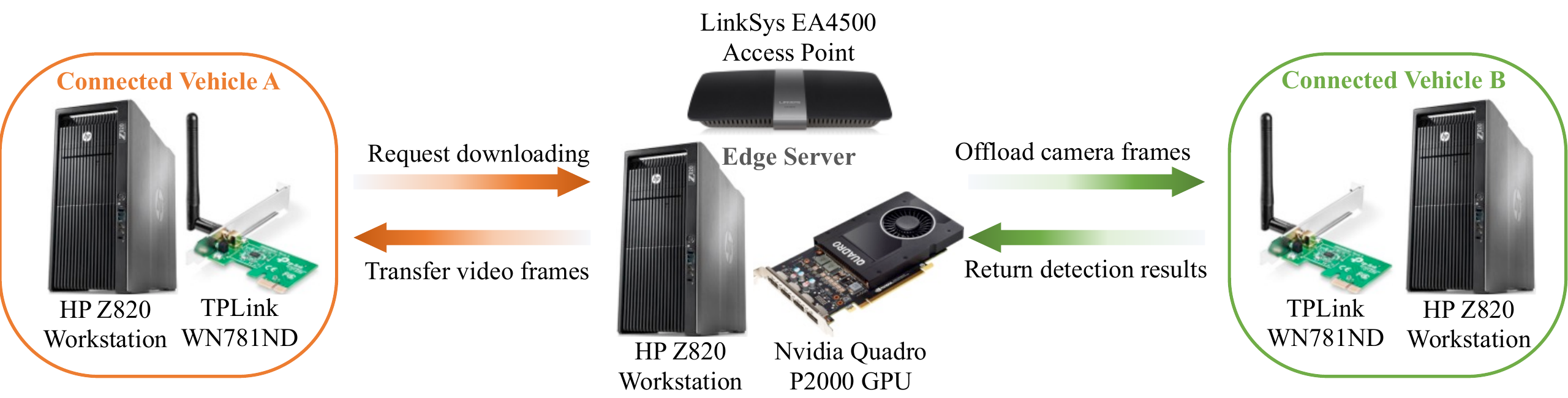}
\caption{\revision{Overview of the testbed implementation for our preliminary experiments. It consists of three components: an edge server and two connected vehicles. The edge server is built upon an HP Z$820$ workstation equipped with a NVIDIA QUADRO P$2000$ GPU and a LINKSYS EA$4500$ AP which is directly connected to the workstation through an Ethernet cable. To eliminate the impact of different wireless network cards on the experiment results, two emulated connected vehicles are deployed with the same TPLINK-WN781ND wireless network cards. Connected vehicle A downloads entertainment video frames from the edge server, while connected vehicle B offloads its camera image frames to the edge server for conducting object detection.}}
\label{fig:fig1_r1}
\end{figure*}

\journal{
Conventional wireless communication networks, for example, the downlink in Wi-Fi and Long-Term Evolution (LTE) networks acquire a higher bandwidth, peak data rate, throughput, and spectral efficiency than their uplink \cite{ott2004drive,huang2012close}. However, most current edge-assisted applications and services, such as high definition (HD) map generation \cite{9963609} and mobile augmented reality \cite{wang2020user} are no longer downlink traffic dominant only. Future MDT-assisted connected services are prone to have more uplink traffic. Therefore, traditional network resource allocation designs that favor the downlink traffic are no longer suitable for supporting QoS/QoE in automotive edge computing networks. In this section,  we conduct experimental measurement studies to explore the wireless network performance changes while heavy uplink and downlink traffic co-exist, and demonstrate the unfairness in traditional IEEE 802.11 wireless networks.
}
\journal{
\begin{figure*}[t]
\centering
\subfigure[]
{\includegraphics[width=0.24\textwidth]{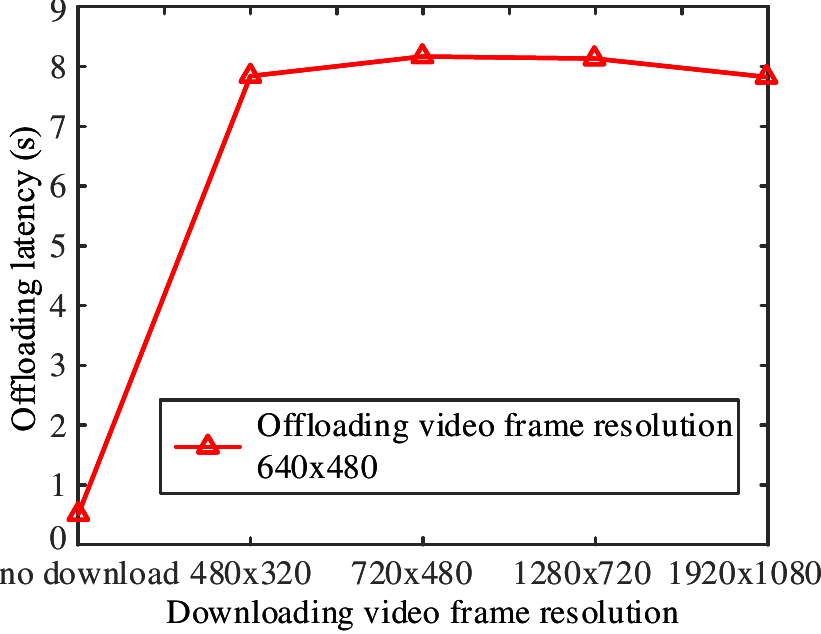}\label{fig:offlatency}}
\subfigure[] 
{\includegraphics[width=0.24\textwidth]{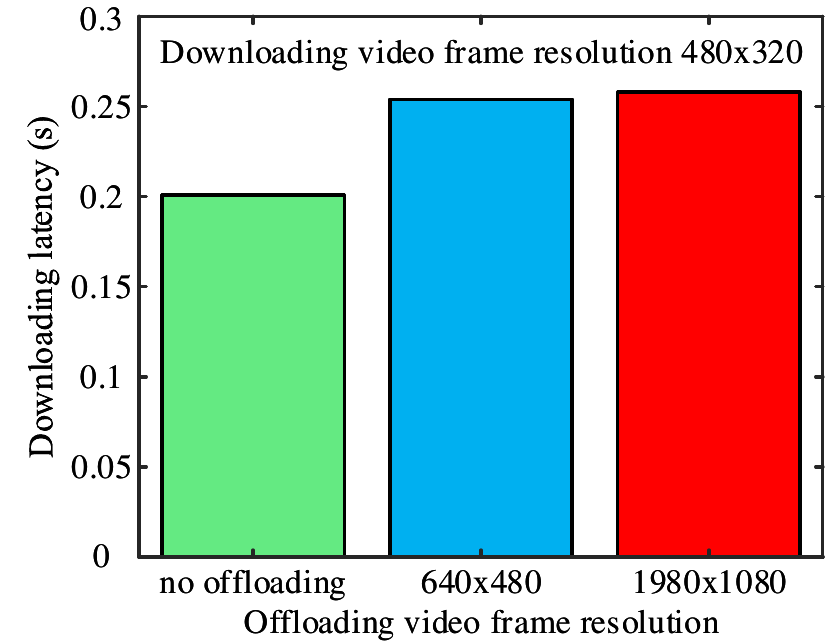}\label{fig:downlatency}} 
\subfigure[]
{\includegraphics[width=0.24\textwidth]{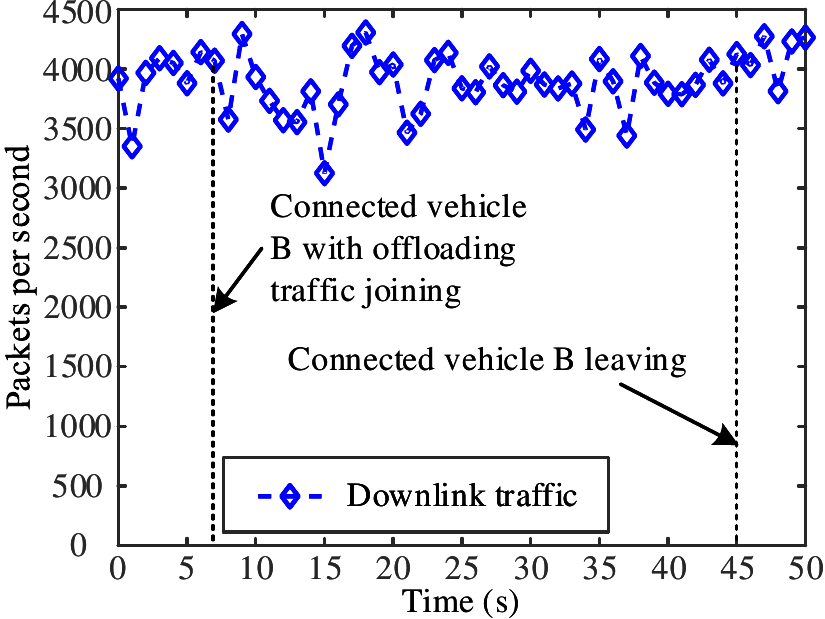}\label{fig:downtra}}
\subfigure[] 
{\includegraphics[width=0.24\textwidth]{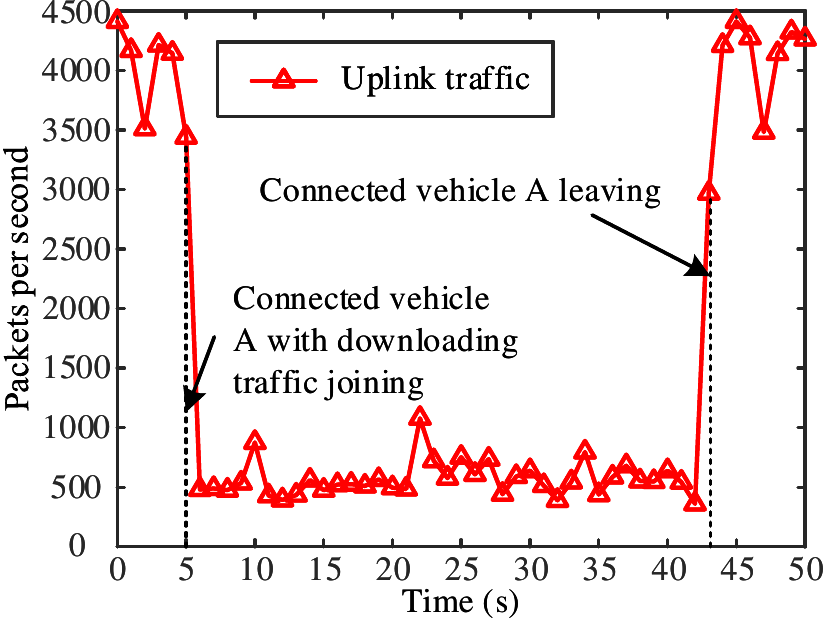}\label{fig:uptra}} 
\caption{\journal{Experimental measurement results. (a) Average offloading latency per frame; (b) Average downloading latency per frame; (c) The number of downlink packets; and (d) The number of uplink packets.}}
\label{fig:traffic}
\end{figure*}

\textbf{Experimental study.} In this paper, we mainly focus on IEEE 802.11 wireless technology, since compared with LTE, IEEE 802.11 is a better communication approach for data transmission between connected vehicles and edge servers, due to its higher data rate and lower monetary cost. In addition, since both heavy uplink traffic (e.g., offloaded sensing data) and downlink traffic (e.g., downloaded entertainment videos) may co-exist in connected vehicle networks, we conduct an experimental measurement study to explore the network performance changes while heavy uplink and downlink traffic co-existing. Our testbed is deployed as shown in Fig. \ref{fig:fig1_r1}. There are one edge server and two emulated connected vehicles in our testbed. The edge server consists of a computation unit and a communication unit. The computation unit is implemented on a HP Z$820$ workstation that is equipped with a NVIDIA QUADRO P$2000$ GPU. The communication unit is a LINKSYS EA$4500$ access point (AP), which is directly connected to the computation unit through a Ethernet cable. In order to eliminate the impact of different wireless network cards on the experiment results, two emulated connected vehicles are equipped with the same TPLINK-WN781ND wireless network cards. In our experiments, connected vehicle A downloads entertainment videos from the edge server, while connected vehicle B offloads its real-time camera captured image frames to the edge server for conducting object detection, which is one of the most important steps of edge-assisted intelligent driving \cite{lin2018architectural}. The object analytics module in the edge server is designed based on the YOLOv{3} framework \cite{yolov3}.

\textbf{Our observations.} We measure the latency of downloading/offloading each video frames, and calculate the average latency. Our observations are described as follows: 
\begin{itemize}
\item As shown in Fig. \ref{fig:offlatency}, the average offloading latency per frame is dramatically increased, approximately $1600$\%, after connected vehicle A with downloading traffic joining the network.
\item As shown in Fig. \ref{fig:downlatency}, the average downloading latency per frame is only increased to approximately $125$\% after connected vehicle B with offloading traffic joining the network, which is much less than the increase of the average offloading latency.
\item As shown in Fig. \ref{fig:offlatency}, when the downloading video frame resolution increases (i.e., increasing the total downloading data size), the increment of the average offloading latency per frame does not change much. The same observation is obtained in the downlink scenario, as shown in Fig. \ref{fig:downlatency}.
\end{itemize}

Then, we conduct another test to explore the reasons of the aforementioned observations. We measure the traffic of downlink and uplink, as depicted in Fig. \ref{fig:downtra} and Fig. \ref{fig:uptra}, respectively. In Fig. \ref{fig:uptra}, we find that a big uplink throughput drop occurs from $5$ sec to $43$ sec, when connected vehicle A with downloading traffic joins the network. However, there is no obvious downlink throughput degrade in Fig. \ref{fig:downtra}, when connected vehicle B with uplink traffic joins the network. This is because usually, AP is set up with a shorter Arbitration Inter-Frame Space Number (AIFSN) value for voice and video traffic (e.g., AIFSN = 1) than its associated stations (e.g., AIFSN = 2) \cite{5929298,NETGEAR, wangrethinking}. Thus, \textit{an AP obtains a higher priority for channel contention than its associated stations (i.e., the downlink traffic has a higher priority than the uplink traffic during contention)}, which significantly impacts the transmission efficiency of connected vehicle B offloading its camera captured video frames. Although, we may set the same AIFS value for both AP and stations in order to eliminate above unfairness, the more complicated and fierce channel contention may still badly influence on the throughput of both downlink and uplink.  

\textbf{Summary.} MDT-assisted connected services with edge computing will suffer from a significant performance degradation due to the network resource allocation unfairness between downlink and uplink in traditional wireless communication networks. \textit{It is imperative to systematically investigate the impartial network resource allocation in automotive edge computing networks towards fast and reliable MDT-assisted connected services.}
}

\section{Proposed \textit{FAIR} Automotive Edge Networking System}
\label{sc:Proposed Scheme}
\begin{figure}[t]
\centering
\includegraphics[width=0.48\textwidth]{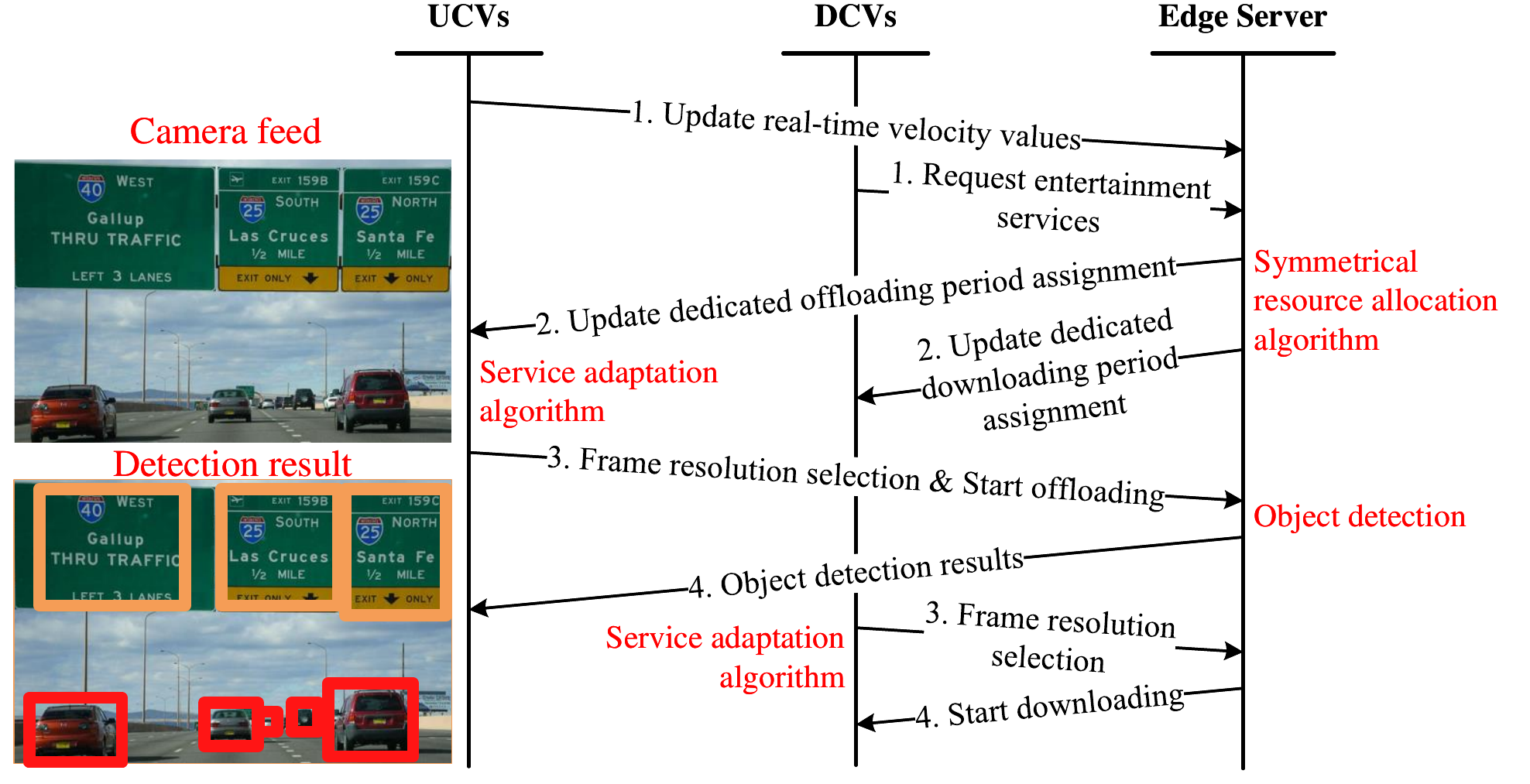}
\caption{\revision{Overview of the proposed \textit{FAIR}. The core of \textit{FAIR} is our proposed symmetrical network resource allocation algorithm deployed at the edge server and proposed service adaptation algorithm equipped on intelligent vehicles. The symmetrical network resource allocation algorithm can mitigate the service performance degradation incurred by the asymmetrical radio resource allocation for uplink and downlink connections. The service adaptation algorithm can dynamically adjust configurations of environmental sensing and frame resolution according to the reserved service period and user preference.}}
\label{fig:Overview}
\end{figure}

\journal{
Based on the above observations, in this section, we propose \textit{FAIR}, an end-to-end automotive edge networking system, which can provide \underline{f}ast, sc\underline{a}lable, and \underline{i}mpa\underline{r}tial connected services for intelligent vehicles in a variety of traffic scenes. Fig. \ref{fig:Overview} illustrates the overview of the proposed \textit{FAIR}. To mitigate the service performance degradation incurred by the asymmetrical radio resource allocation for uplink and downlink connections, we design a new symmetrical network resource allocation algorithm deployed at edge servers to proactively and impartially reserve service periods for individual intelligent vehicles that request dissimilar connected services. In addition, a service adaptation algorithm deployed on intelligent vehicles is proposed to dynamically adjust configurations of environmental sensing and frame resolution according to the reserved service period and user preference.
}

\begin{table}[t]
\centering
\caption{Table of Notation}
\begin{tabular}{|l|l|l|l|l|l|l|}
\hline
Variable         & Description  \\ \hline
$V$   & Speed of a connected vehicle ($m/s$)   \\ \hline
$fps$   & Frame rate (frame/second)   \\ \hline
$\Delta S$   & The area of camera captured different surroundings ($m^2$)   \\ \hline
$k\cdot s$      & Image frame resolution (pixels)   \\\hline
$\gamma$     & The number of bits carried by one pixel (bits)       \\\hline
$R$     & Average wireless data rate       \\\hline
$LU$   & Data offloading latency of an image frame (second)        \\ \hline
$LD$   & Data offloading latency of a video frame (second)        \\ \hline
$EU$   & Per frame energy consumption for offloading (J)       \\ \hline
$ED$   & Per frame energy consumption for downloading  (J)        \\ \hline
$UU$   & Utilization of the assigned $DOP$       \\ \hline
$UD$   & Utilization of the assigned $DDP$        \\ \hline
$EL$   & Life-time energy consumption (J)        \\ \hline
$P_{tr}$   & Average power consumption of transmission phase (W)    \\ \hline
$P_{rev}$   & Average power consumption of receiving data (W)     \\ \hline
\revision{$\mathcal{N}$}   & \revision{The set of all intelligent vehicles}   \\ \hline
\revision{$\mathcal{N}_{d}$}   & \revision{The set of all DCVs}   \\ \hline
\revision{$\mathcal{N}_{u}$}   & \revision{The set of all UCVs}   \\ \hline
\revision{$\theta$}   & \revision{Horizontal field of view of a camera ($\circ$)}   \\ \hline
\revision{$l$}   & \revision{$3$-D measurement range of a camera ($m$)}   \\ \hline
\revision{$\sigma$}  & \revision{Ratio of the areas of UCVs' captured different surroundings}   \\ \hline
\revision{$\omega$}   & \revision{Positive weight parameter customizing user preference}   \\ \hline
\end{tabular}
\label{tb:notation}
\end{table}

\begin{figure}[t]
\centering
\includegraphics[width=0.44\textwidth]{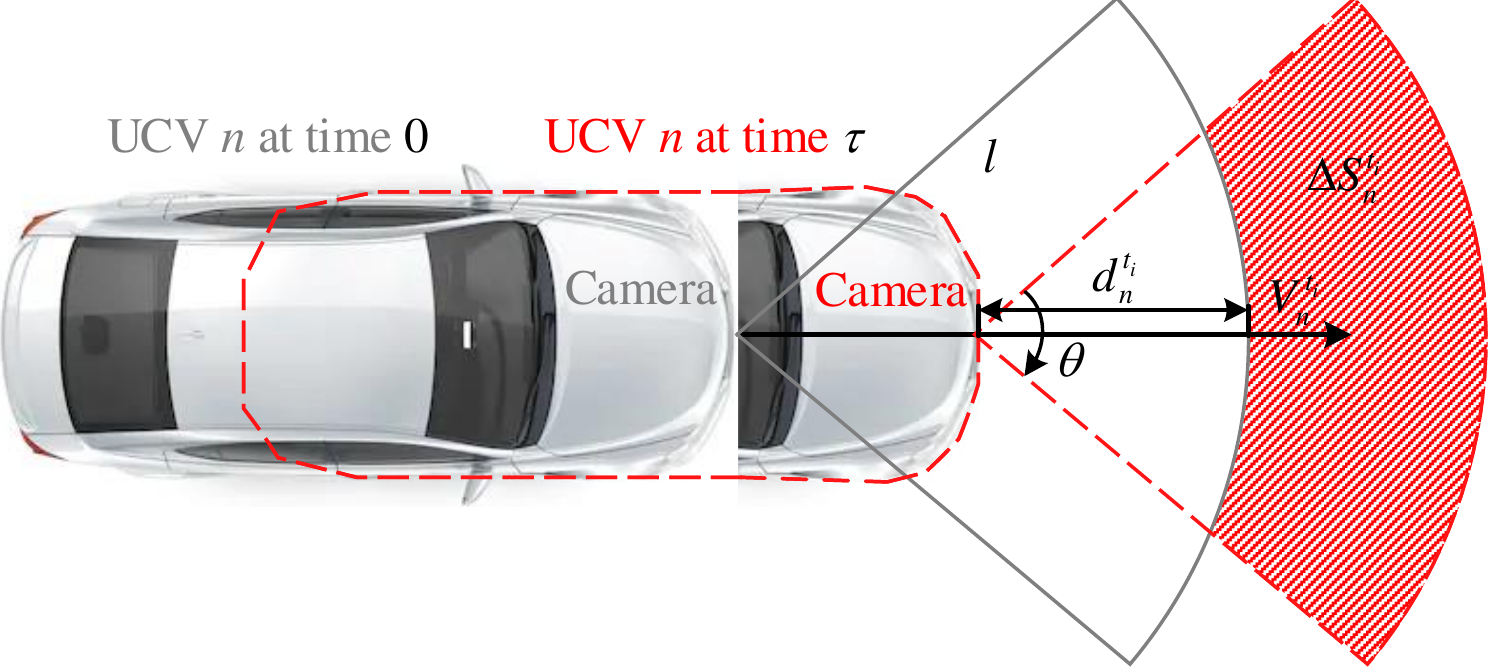}
\caption{The area change of camera captured image frames while moving.}
\label{fig:vehicle_speed}
\end{figure}

\subsection{Key Factors in Edge-assisted Intelligent Vehicle}
\label{ssc:Key factors}
\journal{
We consider an end-to-end automotive edge computing system
consists of a roadside edge server $a$ and $N$ intelligent vehicles that request connected services. Denote $N_{d}$ as the number of intelligent vehicles requesting connected services that downloading traffic are dominant, e.g., streaming entertainment videos. While $N_{u}$ denotes the number of intelligent vehicles requesting connected services that uploading traffic are dominant, e.g., offloading on-board camera captured images to the edge server for further processing). In this paper, the intelligent vehicles with downloading and uploading traffic are named as DCVs and UCVs, respectively. Furthermore, to make our system practical, vehicles are capable of requesting both downloading traffic dominant and uploading traffic dominant services concurrently. Thus, $N_{d}+N_{u}\geq N$. Denote $\mathcal{N}$, $\mathcal{N}_{d}$, and $\mathcal{N}_{u}$ as the set of all intelligent vehicles served by the edge server, DCVs, and UCVs, respectively.

To thoroughly study the correlations between the performance of connected services and the state of vehicles, such as speed and user preference, we define four analytic factors for intelligent vehicles with automotive edge computing.
}

\subsubsection{Velocity of intelligent vehicles} 
\journal{Denote $V_{n}^{t}$ as the real-time velocity of a UCV at time $t$, where $n\in \mathcal{N}_{u}$. Assume each intelligent vehicle is equipped with cameras with $l$ meter $3$-D measurement range and $\theta^\circ$ horizontal field of view. The area of the surrounding environment that the UCV's camera can capture will change while the UCV is moving a time period $\tau$. As illustrated in Fig. \ref{fig:vehicle_speed}, we model the area of the changed surrounding at time $t$ as $\Delta S_{n}^{t}=(2l-\theta\cdot d_{n}^{t})\cdot V_{n}^{t}$, which also describes the area of new captured surroundings by the on-board camera is determined by the real-time velocity of the vehicle. Therefore, based on the above analysis, we define that
\begin{equation}
\label{eq:speed}
    \sigma^{t}_{(n,m)}=\frac{\Delta S_{n}^{t}}{\Delta S_{m}^{t}} = \frac{(2l-\theta\cdot d_{n}^{t})\cdot V_{n}^{t}}{(2l-\theta\cdot d_{m}^{t})\cdot V_{m}^{t}}, n, m \in \mathcal{N}_{u}, 
\end{equation}\noindent
where, $d_{n}^{t}=V_{n}^{t}\cdot \tau$, $d_{m}^{t}=V_{m}^{t}\cdot \tau$.
}

\subsubsection{Frame rate} 
\journal{In this paper, the frame rate, $fps$, is defined as the number of image frames that a DCV downloads for in-vehicle entertainment or a UCV offloads for further processing such as object detection per second. 
\textbf{For UCVs.} When the velocity of a UCV is high, the UCV has to offload more camera captured frames to the edge server for a reliable and safe driving assistance since it has a larger $\Delta S_{n}^{t}$. Assume the minimum acceptable offloading frame rates for UCVs $n$ and $m$, $n,m\in \mathcal{N}_{u}$, are $fps_{n}^{t}$ and $fps_{m}^{t}$, respectively, we can define the correlation between $fps_{n}^{t}$ and $fps_{m}^{t}$ as
\begin{equation}
\label{eq:fps}
    fps_{m}^{t} = \frac{1}{\sigma^{t}_{(n,m)}}\cdot fps_{n}^{t}.
\end{equation}
\textbf{For DCVs.} A more smooth entertainment video experience requires a higher downloading frame rate, which may also requires a higher data rate.}

\subsubsection{Frame resolution} \journal{Frame resolution is crucial for the performance of both entertainment service and edge-assisted driving assistance.
\textbf{For UCVs.} In this paper, we use $k_{n}^{f_{u}}\cdot s_{n}^{f_{u}}$ (pixels) to represent the frame resolution of the $f_{u}$th image captured by the vehicle's on-board camera. Denote $\gamma$ as the color depth of a frame, i.e., the number of bits required to represent the information carried by one pixel. The data size of the camera captured image $f_{u}$ at time $t$ is calculated as $k_{n}^{f_{u}}\cdot s_{n}^{f_{u}}\cdot\gamma$ bits. We model the offloading latency of the frame $f_{u}$ as 
\begin{equation}
\label{equ:offloading latency}
    LU_{n}^{f_{u}} = \frac{k_{n}^{f_{u}}\cdot s_{n}^{f_{u}}\cdot\gamma}{R_{n}^{t}},
\end{equation}\noindent
where $R_{n}^{t}$ is the average transmission data rate of UCV $n$ at time $t$. Thus, the image offloading latency is a function of wireless data rate and image resolution. The image resolution can be adapted to meet the latency requirement under wireless network dynamics. For example, the UCV may proactively degrade its camera image resolution when the wireless data rate is low. However, the connected service performance, e.g., object detection accuracy, may be affected by the image resolution as well. A lower image resolution usually leads to a worse mean average precision of an object detection \cite{huang2017speed}. Therefore, it is critical for UCVs with high velocity to keep a high camera image resolution for a better service performance and a safer driving. Furthermore, as the data size of object detection results is usually small, we do not consider the latency caused by returning the results in this work.

\textbf{For DCVs.} Consider a DCV $g\in \mathcal{N}_{d}$ that request downloading entertainment videos while moving. Let the frame resolution of the requested video at time $t$ is $k_{g}^{f_{d}}\cdot s_{g}^{f_{d}}$ pixels. We model the downloading latency as
\begin{equation}
\label{equ:downloading latency}
    LD_{g}^{f_{d}} = \frac{k_{g}^{f_{d}}\cdot s_{g}^{f_{d}}\cdot\gamma}{R_{g}^{t}},
\end{equation}\noindent
where $R_{g}^{t}$ is the wireless data rate of DCV $g$ at time $t$.
}

\subsubsection{Energy consumption of connected vehicles} 

\begin{figure}[t]
\centering
\includegraphics[width=0.44\textwidth]{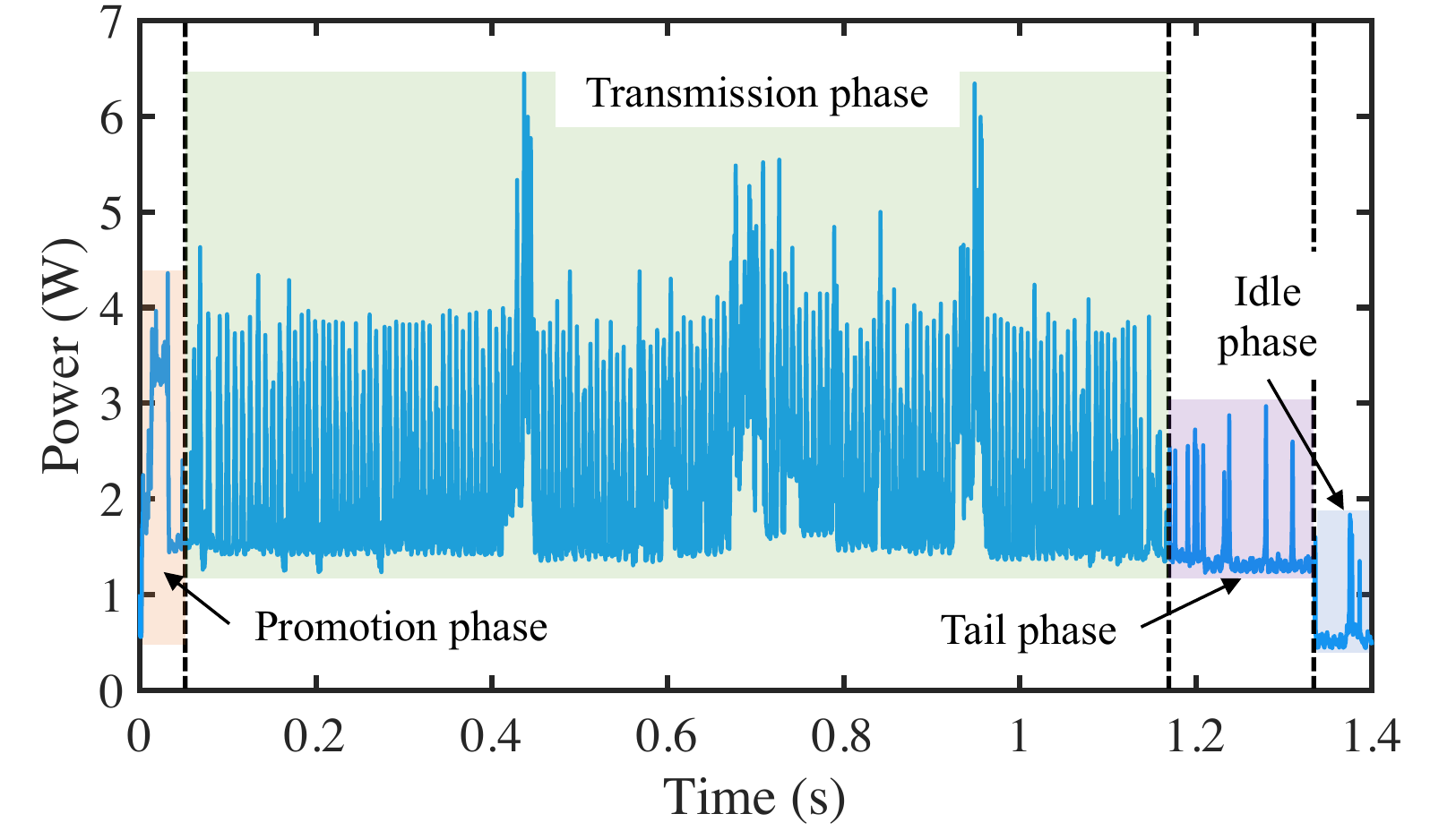}
\caption{\journal{Power consumption of promotion, transmission, tail, and idle phases.}}
\label{fig:transphase}
\end{figure}

{\journal{\textbf{For UCVs.} To estimate the energy efficiency of performing the edge-assisted intelligent driving, in particular, camera sensing and image offloading, we propose \textit{per frame energy consumption}, a new performance metric that appraises the total amount of energy consumed in a connected vehicle by performing the edge-assisted intelligent driving on one image frame.
In this work, the per frame energy consumption of image frame $f_{u}$ in UCV $n\in \mathcal{N}_{u}$ can be defined as

\begin{equation}
    EU^{f_{u}}_{n} = EU^{(f_{u}, n)}_{cam} + EU^{(f_{u}, n)}_{tr},
\end{equation}\noindent
where $EU_{cam}$ and $EU_{tr}$ denote the energy consumed by sampling an image and transmitting the image frame to a edge server, respectively. Image sampling is the process of transferring the vehicle's camera sensed continuous light signal to a processable digital image frame. $EU_{cam}$ is a function of image resolution $k \cdot s$, e.g., a larger $k \cdot s$ will result in a higher $EU_{cam}$. Furthermore, since the energy consumed by image sampling contributes the largest portion of the per frame energy consumption \cite{wang2020energy}, the camera sampling frequency should be adaptable to improve the energy efficiency. Therefore, we propose to co-adapt the image sampling and offloading frequency.  

During the transmission of a single image frame in wireless communication networks, e.g., Wi-Fi, the vehicle's wireless interface undergoes four phases: promotion, data transmission, tail, and idle. As depicted in Fig. \ref{fig:transphase} (power consumption measured in our testbed), the wireless interface first steps into the promotion phase once a transmission request is initiated. Then, the image data is sent to the edge server when the data transmission phase is invoked. The completion of transmission triggers the tail phase, where the wireless interface is compelled to stay active for a fixed duration and to wait for either a new transmission request or service feedback from the edge server. The wireless interface finally enters the idle phase to save energy if there is no new request initiated or feedback received. We model the energy consumption of transmitting image frame $f_{u}$ in UCV $n\in \mathcal{N}_{u}$ as

\begin{equation}
\label{eq:EU}
    EU^{(f_{u}, n)}_{tr} = e_{pro} + P^{(n,f_{u})}_{tr}\cdot LU_{n}^{f_{u}} + e_{tail},
\end{equation}
\noindent
where $P^{(n,f_{u})}_{tr}$ is the average power consumption of the data transmission phase; $e_{pro}$ and $e_{tail}$ are the energy consumed in promotion and tail phases, respectively. $e_{pro}$ and $e_{tail}$ are constant.

\textbf{For DCVs.} The per frame energy consumption of downloading video frame $f_{d}$ in DCV $g\in \mathcal{N}_{d}$ can be modeled as

\begin{equation}
ED_{g}^{f_{d}} = P^{(g,f_{d})}_{rev}\cdot LD_{g}^{f_{d}},
\end{equation}\noindent
where $P^{(g,f_{d})}_{rev}$ is the average power consumption of receiving data.

In addition to the per frame energy consumption, we propose a new performance metric, \textit{life-time energy consumption}, to estimate the total amount of energy consumed by performing uploading and downloading connected services for a certain amount of time $\Delta T$, which is defined as

\begin{equation}
    EL(\Delta T) = \sum^{f_{u}} EU + \sum^{f_{d}} ED.
\end{equation}

}}


\subsection{Symmetrical Resource Allocation Algorithm}
\label{ssc:symmetrical allocation}
\journal{As discussed in Section \ref{ssc:Key factors}, to assure a reliable and safe edge-assisted driving assistance, the amount of offloading camera image data should be proportional to the velocity of the UCV. In this work, the core of our proposed end-to-end automotive edge networking system, \textit{FAIR}, is that \textit{the edge server proactively and periodically reserves a \underline{d}edicated \underline{o}ffloading \underline{p}eriod (DOP) for each served UCV based on its real-time velocity.} In particular, the value of the reserved DOP at time $t$ for UCV $n$ is with respect to the real-time velocity $V_{n}^{t}$. Thus, we define DOP as $DOP_{n}=\psi(V_{n}^{t})$, where a larger DOP will be assigned to the UCV with a higher velocity for supporting an edge-assisted driving assistance with high frame rate and resolution. Furthermore, to make the reserved DOPs can be adapted to the environment dynamics, including the varying velocity and position of intelligent vehicles, the edge server will update the DOP reservation once every $T$ s. After completing the DOP reservation for all UCVs, the edge server will allocate a \underline{d}edicated \underline{d}ownloading \underline{p}eriod (DDP) for each DCVs. Since the required data volume of the in-vehicle entertainment is usually irrelevant to vehicle's velocity, we design that the length of the DDP assigned for each DCV is the same, where
\begin{equation}
\label{eq:DDP}
    DDP_{g} = \frac{T-\sum_{n\in\mathcal{N}_{u}}(DOP_{n})}{N_{d}}, g\in \mathcal{N}_{d}.
\end{equation}

}

\begin{algorithm}[t]
\DontPrintSemicolon
\footnotesize
\SetNoFillComment
\label{alg:1}
\caption{\textit{FAIR} Symmetrical resource allocation algorithm}
\KwIn{Vector of UCV speed: $\vec{v}$; the number of UCVs in cases 1, 2, 3: $\kappa_{1}$, $\kappa_{2}$, $\kappa_{3}$; data rate: $R$; total number of UCVs and DCVs: $N_{u}$, $N_{d}$; period of resource allocation: $T$; acceptable frame rate: $fps_{0}$; highest camera frame resolution: $k_{max}\cdot s_{max}$.}
\KwOut{Sets of RSU server allocated DOPs and DDPs: $\overrightarrow{DOP}$ and $\overrightarrow{DDP}$.}
$T_{0}\gets T$;\;
\If{$\kappa_{1}> 0$}{
Find $R_{min}$; \Comment{Case 1 triggered}\\
$DOP_{max}\gets$ calculate via (12); \\
$T_{0}\gets T_{0} - \sum DOP_{max}$;\\
}
\Else{
$DOP_{max}\gets$ calculate via (10);\\
$T_{0}\gets T_{0} - DOP_{max}$;\\
}

\If{$\kappa_{2}>0$}{
    $DOP_{n\in \mathcal{K}_{2}}\gets$ calculate via (11); \Comment{Case 2 triggered}\\ 
    $T_{0}\gets T_{0}-\sum^{\kappa_{2}}{DOP_{n\in\mathcal{K}_{2}}}$;
    }
    \For{$n \in \mathcal{N}_{u}$}{
    $DOP_{n}\gets$ calculate via (11);\\
    \If{$\kappa_{3}>0$}{
    $DOP_{n\in\mathcal{K}_{3}}\gets$ calculate via (13); \Comment{Case 3 triggered}\\
    $T_{0}\gets T_{0}-\sum_{1}^{\kappa_{3}}{DOP_{n\in\mathcal{K}_{3}}}$;
    }
    \If{$T_{0}\leq DOP_{n}$}{
    $DOP_{n}\gets T_{0}$; \\
    $T_{0}=0$;\\
    \textbf{break};\\}
    $T_{0}\gets T_{0}-DOP_{n}$;\\
    }
\If{$T_{0}>0$}{
    \For{$g=1$ \bf{to} $N_{d}$}{
    $DDP_{g}\gets$ calculate via (9);\Comment{Allocate DDP for DCVs}\\
}}
$\overrightarrow{DOP}\gets DOP_{n}$; $\overrightarrow{DDP}\gets DDP_{g}$;\; 
\Return $\overrightarrow{DOP}$ and $\overrightarrow{DDP}$.
\end{algorithm}

\journal{We assume the velocity of each intelligent vehicle is constant in $T$ s, and define the velocity vector of UCVs as $\vec{v} = \{v^{x}_1,\ldots,v^{y}_{N_{u}}\}$, a $N_{u}$-tuple containing the real-time velocity of each UCV within $T$ s. UCVs in the velocity vector are sorted in descending order in terms of the value of velocity. Denote $\overrightarrow{DOP}$ as the vector of edge reserved $DOP$s for UCVs in $\mathcal{N}_{u}$ within $T$ s. The edge server will reserve the $DOP$ for each UCV every $T$ s by leveraging the updated $\vec{v}$. In our proposed symmetrical wireless resource allocation algorithm, the UCV with the highest velocity will be always reserved with the highest priority. In particular, the UCV with the highest velocity in $T$ obtains the best edge networking resource including the largest reserved $DOP$, where $DOP =DOP_{max}$; the highest camera frame resolution, where $k \cdot s = k_{max}\cdot s_{max}$, with an acceptable frame rate $fps_{0}$. The $DOP_{max}$ is defined as 
\begin{equation}
\label{eq:dopmax allocation}
    DOP_{max} = DOP_{x}=\frac{k_{max}\cdot s_{max}\cdot \gamma}{R_{n}^{t}}\cdot fps_{0}\cdot T.
\end{equation}
Furthermore, $DOP$ of the rest of UCVs will be assigned in sequence according to $\vec{v}$. The reserved $DOP$ of UCV $n$ is defined as
\begin{equation}
\label{eq:dop allocation}
    DOP_{n} = \psi(V_{n}) = \frac{1}{\sigma^{t}_{(x,n)}}\cdot DOP_{max}, n\in \mathcal{N}_{u}.
\end{equation}

Based on the designed principle of service allocation in edge networking systems, we can see that a UCV with a higher real-time velocity can always be assured of a more reliable connected service by reserving a larger $DOP$. In addition, we classify three cases to describe special traffic scenarios, defined as follows.
\begin{enumerate}
    \item Highway. We set $V_{highway}$ as the trigger to determine if the traffic scene is driving on a highway. For example, in the United States of America, the value of $V_{highway}$ is usually over $60$ miles per hour or $26.8$ m/s. To ensure the reliability of the service connection and the safety of driving, our proposed symmetrical service allocation algorithm assigns a $DOP_{max}$ for every UCV $n$ that is classified as driving on highway (i.e., $n\in\mathcal{K}_{1}$). In this case, $DOP_{max}$ is defined as 
    \begin{equation}
    \label{eq:highway allocation}
        DOP^{n\in\mathcal{K}_{1}}_{max} = \frac{k_{max}\cdot s_{max}\cdot \gamma}{R^{t}_{n}}\cdot fps_{0}\cdot T.
    \end{equation}\noindent
    The assigned $DOP_{max}$ aims to assure each UCV is capable of offloading camera captured images with the highest frame resolution $k_{max}\cdot s_{max}$ and the acceptable frame rate $fps_{0}$.
    
    \item Continuous unallocated. UCVs that are not assigned for a $DOP$ in the last $\beta\cdot T$ s. $\kappa_{2}$ and $\mathcal{K}_{2}$ represent the number of UCVs and the set of UCVs in case $2$, respectively. UCVs in this case are provided with the highest $DOP$ allocation priority except the UCVs in case $1$.
    
    \item Temporary stop. This case includes UCVs with a temporary stop incurred by a red traffic light, traffic congestion, or temporary parking. $\kappa_{3}$ and $\mathcal{K}_{3}$ represent the number of UCVs and the set of UCVs in case $3$, respectively. UCVs with a temporary stop only need to opportunistically sense and transmit a camera image to substantiate whether the preceding car's status has changed or not, or if the traffic signal light turns green. Therefore, the lowest priority is assigned to the UCVs in this case, where
    \begin{equation}
    \label{eq:stop allocation}
        DOP_{n\in\mathcal{K}_{2}} = \frac{k_{min}\cdot s_{min}\cdot \gamma}{R_{n}^{t}}.
    \end{equation}
\end{enumerate}

In addition, $DDP$ reserved for each DCV is defined by (\ref{eq:DDP}). Denote $\overrightarrow{DDP}$ as the vector of edge reserved $DDP$s for DCVs in $\mathcal{N}_{d}$ within $T$ s. Algorithm \ref{alg:1} presents the proposed symmetrical resource allocation algorithm in detail.
}

\subsection{Service Adaptation Algorithm}
\label{ssc:adaptation}
\begin{figure}[t]
\centering
\subfigure[EDCA channel access]
{\includegraphics[width=0.24\textwidth]{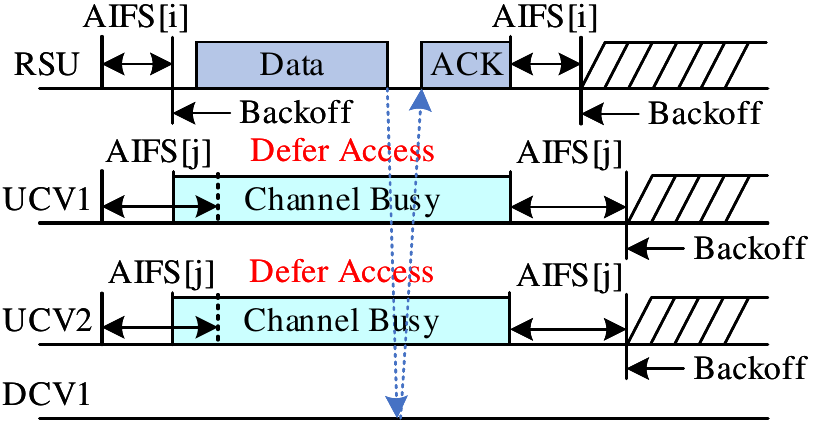}\label{fig:EDCA}}
\subfigure[FAIR channel access] 
{\includegraphics[width=0.24\textwidth]{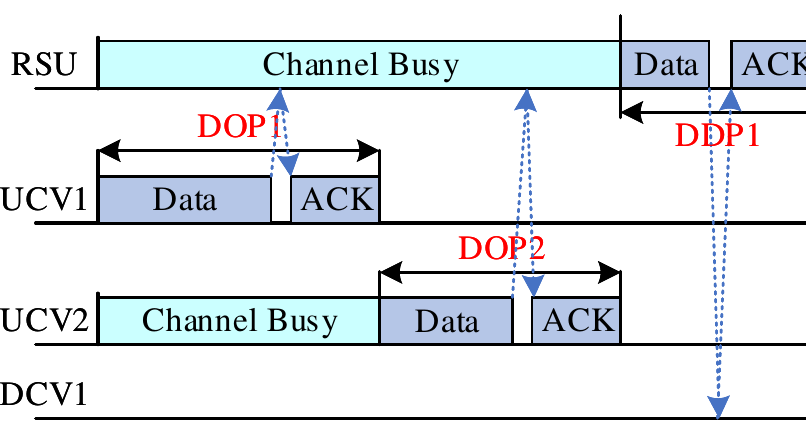}\label{fig:E-Autoaccess}} 
\caption{Comparison of channel access.}
\label{fig:channel access}   
\end{figure}

\journal{
Our proposed symmetrical network resource allocation algorithm proactively and impartially allocates a $DOP$ or a $DDP$ for each connected vehicle that initiates either uploading or downloading traffic. Then, connected vehicles will perform their request connected services through exploiting these allocated $DOP$s and $DDP$s. Therefore, it is imperative to deploy a mechanism on connected vehicles that can dynamically adapt image sensing and offloading (e.g., image frame resolution) for UCVs or downloaded video quality for DCVs (e.g., video frame resolution). To achieve effective adaptations, we propose a service adaptation algorithm to coordinate with the symmetrical resource allocation algorithm. 

As we proposed in the symmetrical resource allocation algorithm, \textit{FAIR} ensures that the UCV with the highest instantaneous speed or UCVs with instantaneous speed over $V_{highway}$ are reserved with $DOP$s that allow the camera image sampling and transmission with the highest frame resolution $k_{max}\cdot s_{max}$. While the image frame resolution of other UCVs are selected in the range of $k_{min}\cdot s_{min}$ to $k_{max}\cdot s_{max}$ based on the optimization problem $\mathscr{P}_{0}$, which aims to achieve an optimal image frame resolution $k^{f_{u}}\cdot s^{f_{u}}$ to balance the offloading energy consumption and the utilization of the assigned $DOP$. $\mathscr{P}_{0}$ is formulated as follows.

\begin{small}
\begin{equation}
\begin{aligned}
\mathscr{P}_{0}: \min_{\{k^{f_{u}},s^{f_{u}}\}} \quad & 
Q_{u} = \omega_{1}\cdot EU^{f_{u}} - \omega_{2}\cdot UU^{f_{u}}\\
s.t. \quad 
    & C_{1}: k^{f_{u}}\cdot s^{f_{u}}\in \{k_{min}\cdot s_{min}, ..., k_{max}\cdot s_{max}\};\\
    & C_{2}: UU^{f_{u}} \leq 1;\\
\end{aligned}
\end{equation}
\end{small}\noindent
where $UU^{f_{u}} = \frac{k^{f_{u}}\cdot s^{f_{u}}\cdot \gamma}{DOP\cdot R^{t}}$ describes the utilization of the assigned $DOP$; and $\omega_{1}$ and $\omega_{2}$ are two positive weight parameters that are introduced to customize the user preference or to discriminate the configuration of different connected services. For example, given a larger $\omega_{1}$ and a smaller $\omega_{2}$, the frame resolution selection strategy will tend to be energy saving, and vice versa. $\omega_{1}$ and $\omega_{2}$ can be specified by either the user or connected service provider. Similarly, frame resolution determination in DCVs is formulated and resolved by

\begin{small}
\begin{equation}
\begin{aligned}
\label{eq:Q1}
\mathscr{P}_{1}: \min_{\{k^{f_{d}},s^{f_{d}}\}} \quad & 
Q_{d} = \Bar{\omega_{1}}\cdot ED^{f_{d}} - \Bar{\omega_{2}}\cdot UD^{f_{d}}\\
s.t. \quad 
    & C_{1}: k^{f_{d}}\cdot s^{f_{d}}\in \{\Bar{k_{min}}\cdot \Bar{s_{min}}, ..., \Bar{k_{max}}\cdot \Bar{s_{max}}\};\\
    & C_{2}: UD^{f_{d}} \leq 1;\\
\end{aligned}
\end{equation}
\end{small}\noindent
where $UD^{f_{d}} = \frac{k^{f_{d}}\cdot s^{f_{d}}\cdot \gamma}{DDP\cdot R^{t}}$ describes the utilization of the assigned $DDP$. Given the above discussion and the formulated problems $\mathscr{P}_{0}$ and $\mathscr{P}_{1}$, we design the service adaptation algorithm, where the pseudo code is delineated in Algorithm \ref{alg:2}.

\begin{algorithm}[t]
\DontPrintSemicolon
\footnotesize
\SetNoFillComment
\label{alg:2}
\caption{\textit{FAIR} Service adaptation algorithm}
\KwIn{$DOP$; $DDP$; $DOP_{max}$; user preference of UCVs: $\omega_{1}$, $\omega_{2}$; user preference of DCVs:$\bar{\omega_{1}}$, $\bar{\omega_{2}}$; $k_{max}\cdot s_{max}$, $k_{min}\cdot s_{min}$; $\bar{k_{max}}\cdot \bar{s_{max}}$, $\bar{k_{min}}\cdot \bar{s_{min}}$; and $R^{t}$.}
\KwOut{Optimal camera frame resolution: $k^{f_{u}}\cdot s^{f_{u}}$ and optimal downloading video resolution: $k^{f_{d}}\cdot s^{f_{d}}$.}
\If{UCV $n\in \mathcal{N}_{u}$}{
\If{$DOP = DOP_{max}$}{
$k^{f_{u}}\cdot s^{f_{u}} \gets k_{max}\cdot s_{max}$;\\
}
\Else{
$k^{f_{u}}\cdot s^{f_{u}} \gets$ solving $\mathscr{P}_{0}$ with $DOP$,  $\omega_{1}$, $\omega_{2}$,\\ $k_{max}\cdot s_{max}$, $k_{min}\cdot s_{min}$, and $R^{t}$;}
}
\ElseIf{DCV $g\in \mathcal{N}_{d}$}{
$k^{f_{d}}\cdot s^{f_{d}} \gets$ solving $\mathscr{P}_{1}$ with $DDP$, $\bar{\omega_{1}}$, $\bar{\omega_{2}}$,\\ $\bar{k_{max}}\cdot \bar{s_{max}}$, $\bar{k_{min}}\cdot \bar{s_{min}}$, and $R^{t}$;
}
\Return $k^{f_{u}}\cdot s^{f_{u}}$ and $k^{f_{d}}\cdot s^{f_{d}}$.
\end{algorithm}

In summary, our proposed \textit{FAIR} automotive networking system not only assures the impartial and proactive resource allocation for both UCVs and DCVs within variant environmental conditions, but also enhances both the energy efficiency of performing connected services and network resource utilization compared to traditional wireless networks, IEEE 802.11 Enhanced Distributed Channel Access (EDCA), as illustrated in Fig. \ref{fig:channel access}.
}

\section{Performance Evaluation}
\label{sc:evaluation}

\begin{figure*}[t]
\centering
\subfigure[]
{\includegraphics[width=0.5\textwidth]{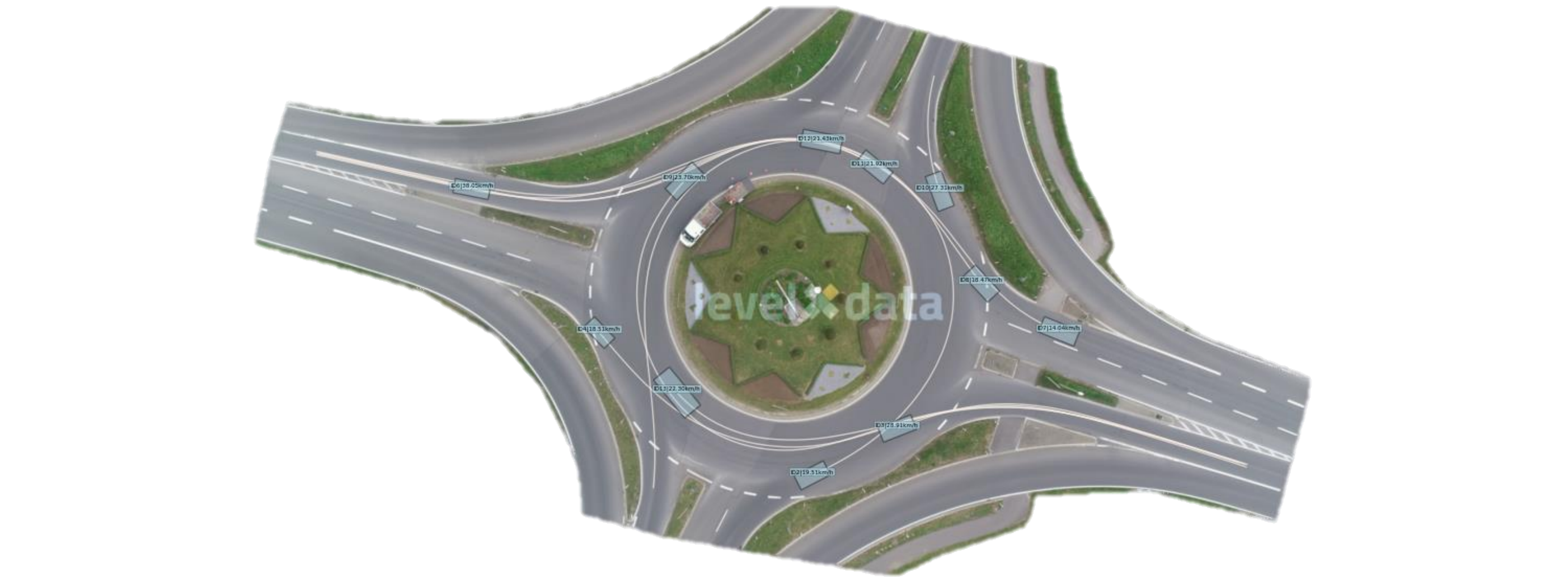}\label{fig:roundmap}}
\subfigure[]
{\includegraphics[width=0.32\textwidth]{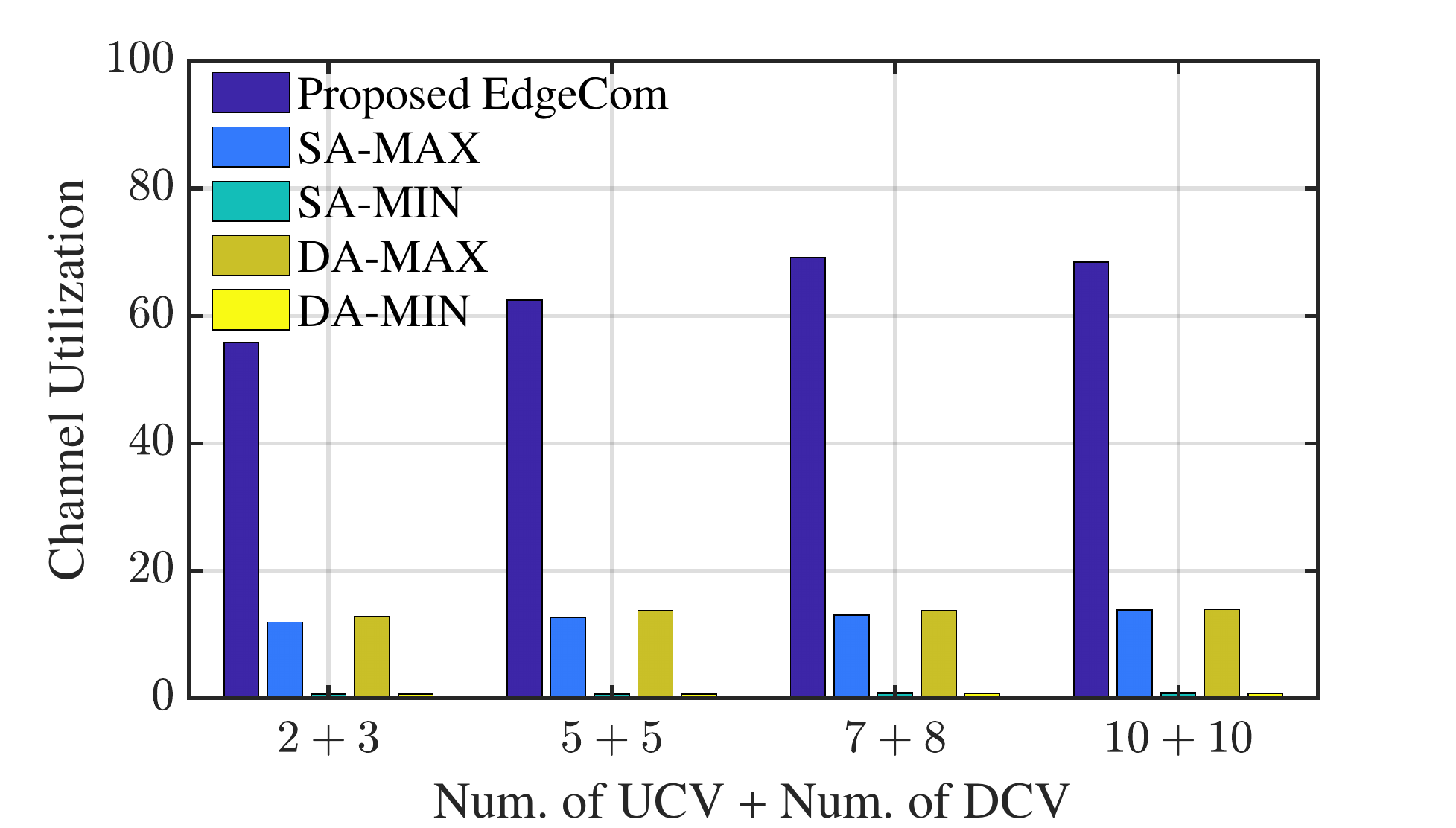}\label{fig:roundUti}}
\subfigure[]
{\includegraphics[width=0.32\textwidth]{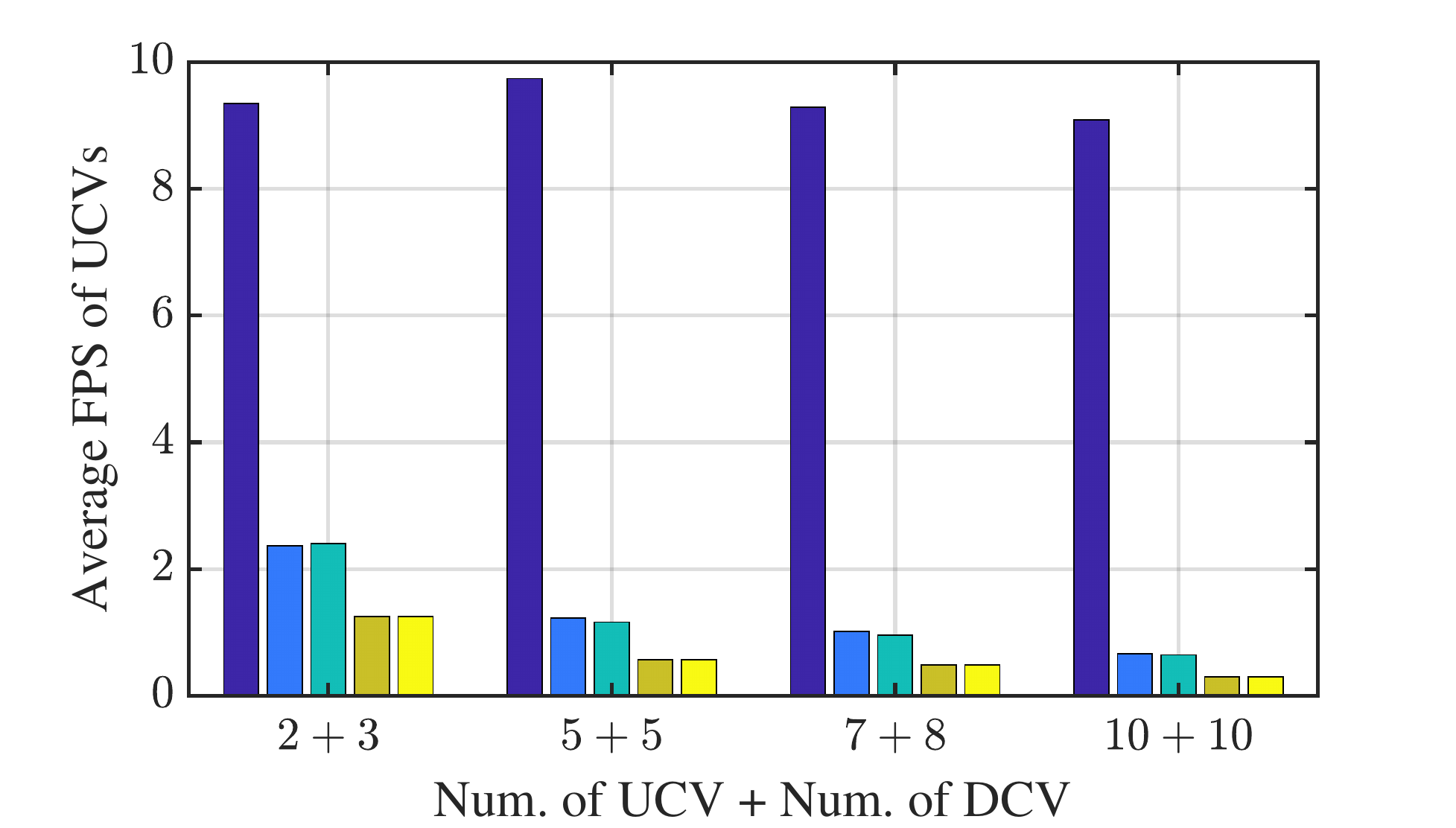}\label{fig:roundfpsu}}
\subfigure[]
{\includegraphics[width=0.32\textwidth]{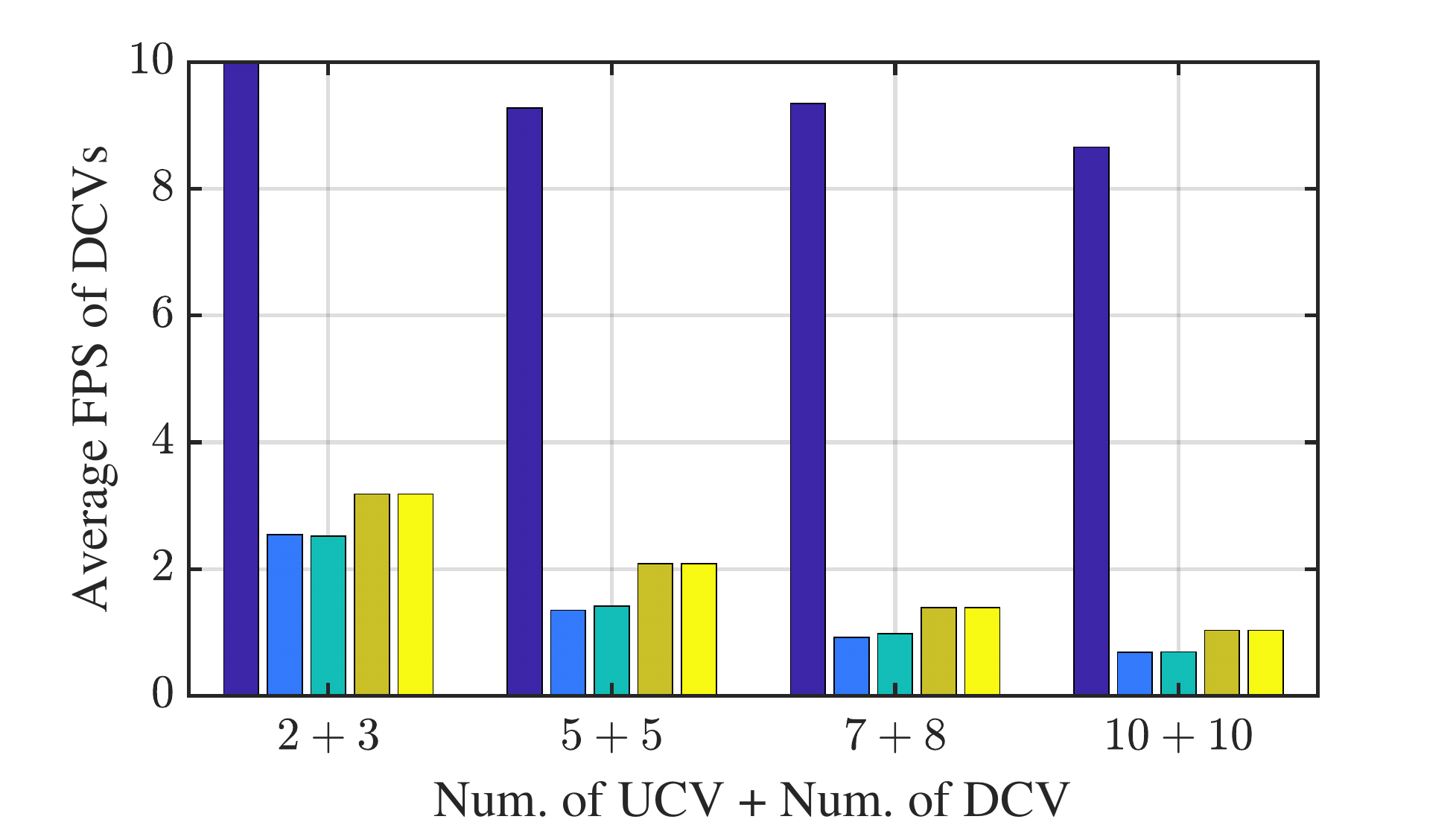}\label{fig:roundfpsd}}
\subfigure[]
{\includegraphics[width=0.32\textwidth]{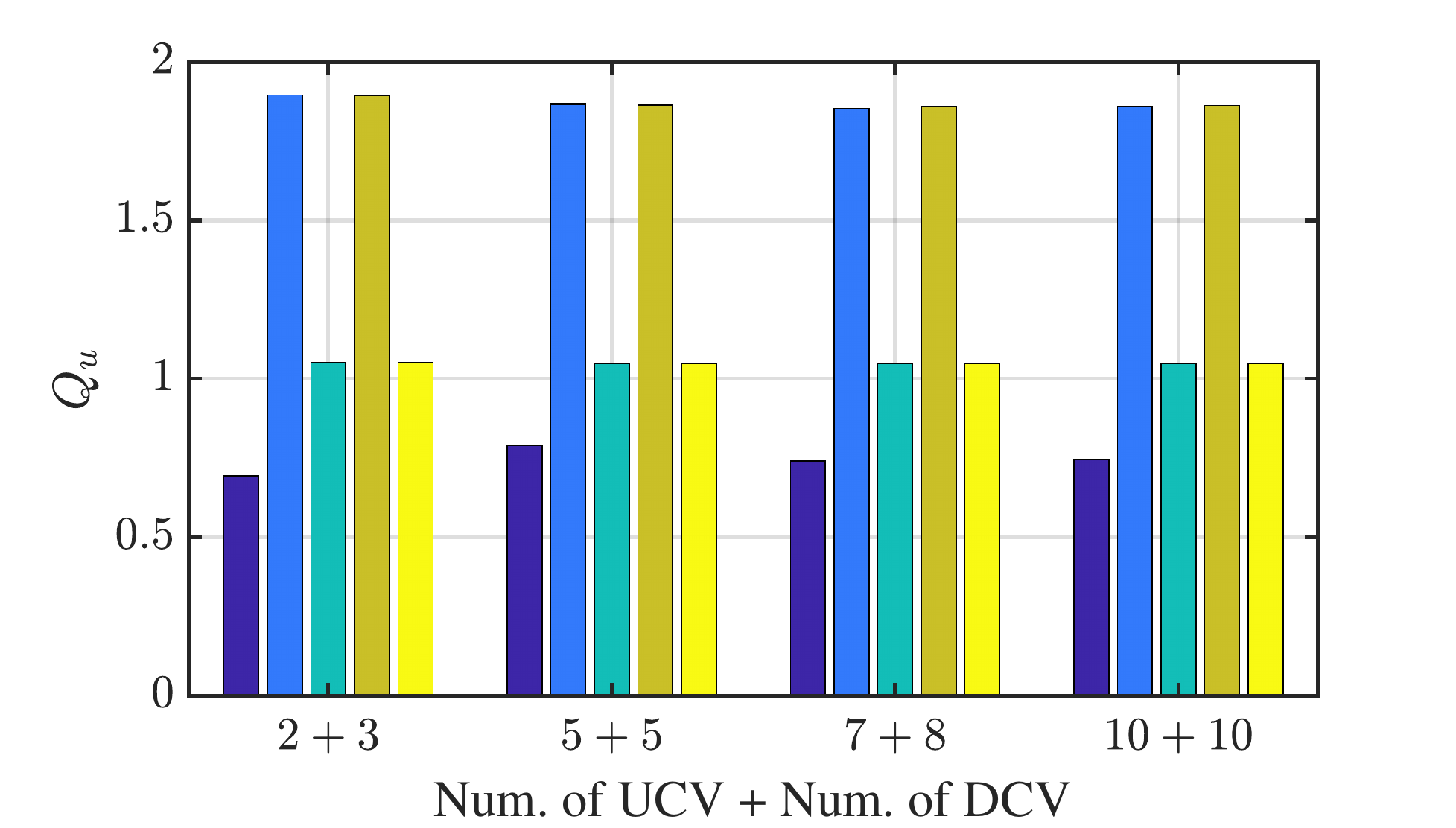}\label{fig:roundqunum}}
\subfigure[]
{\includegraphics[width=0.32\textwidth]{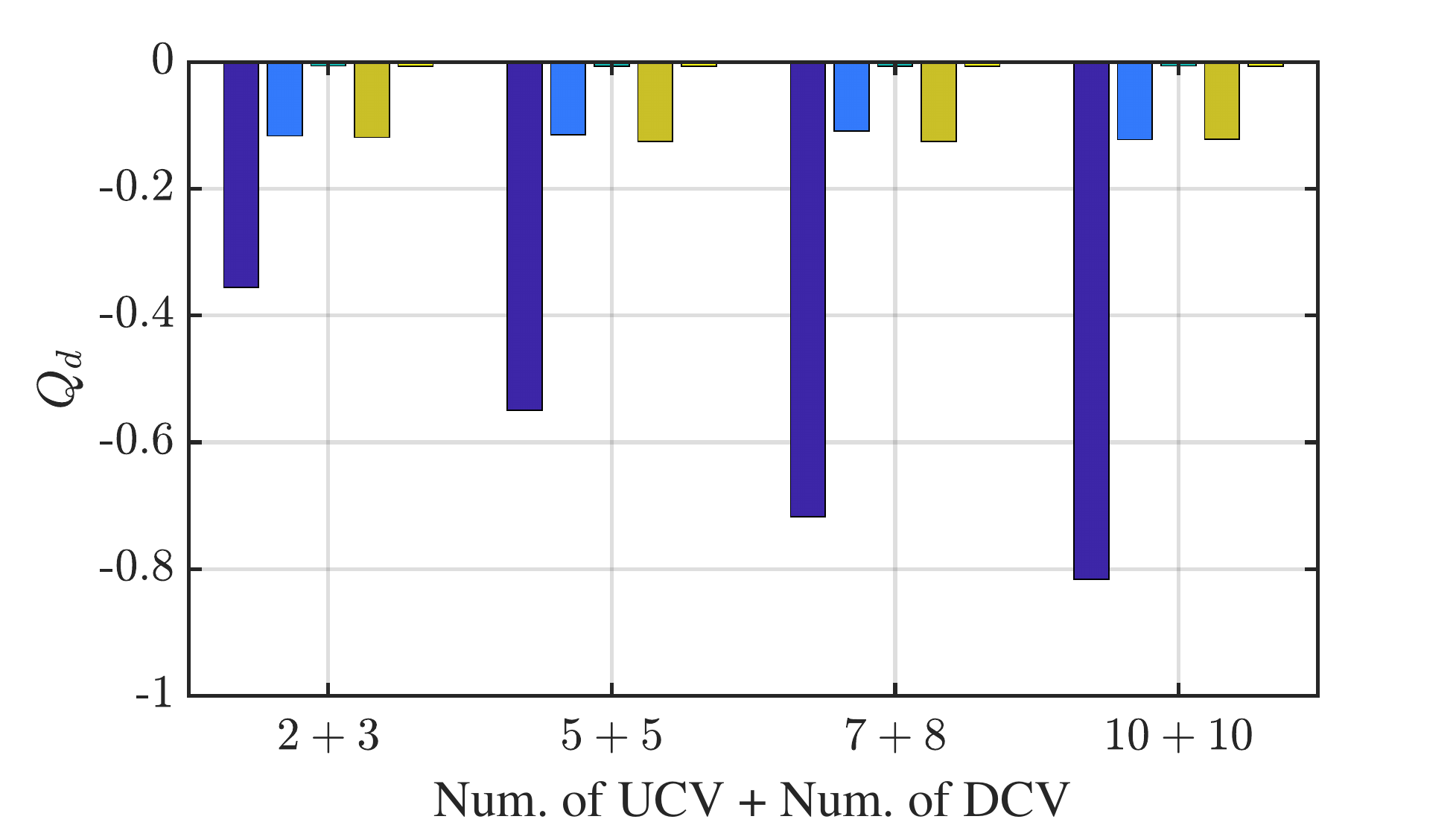}\label{fig:roundqdnum}}
\subfigure[]
{\includegraphics[width=0.32\textwidth]{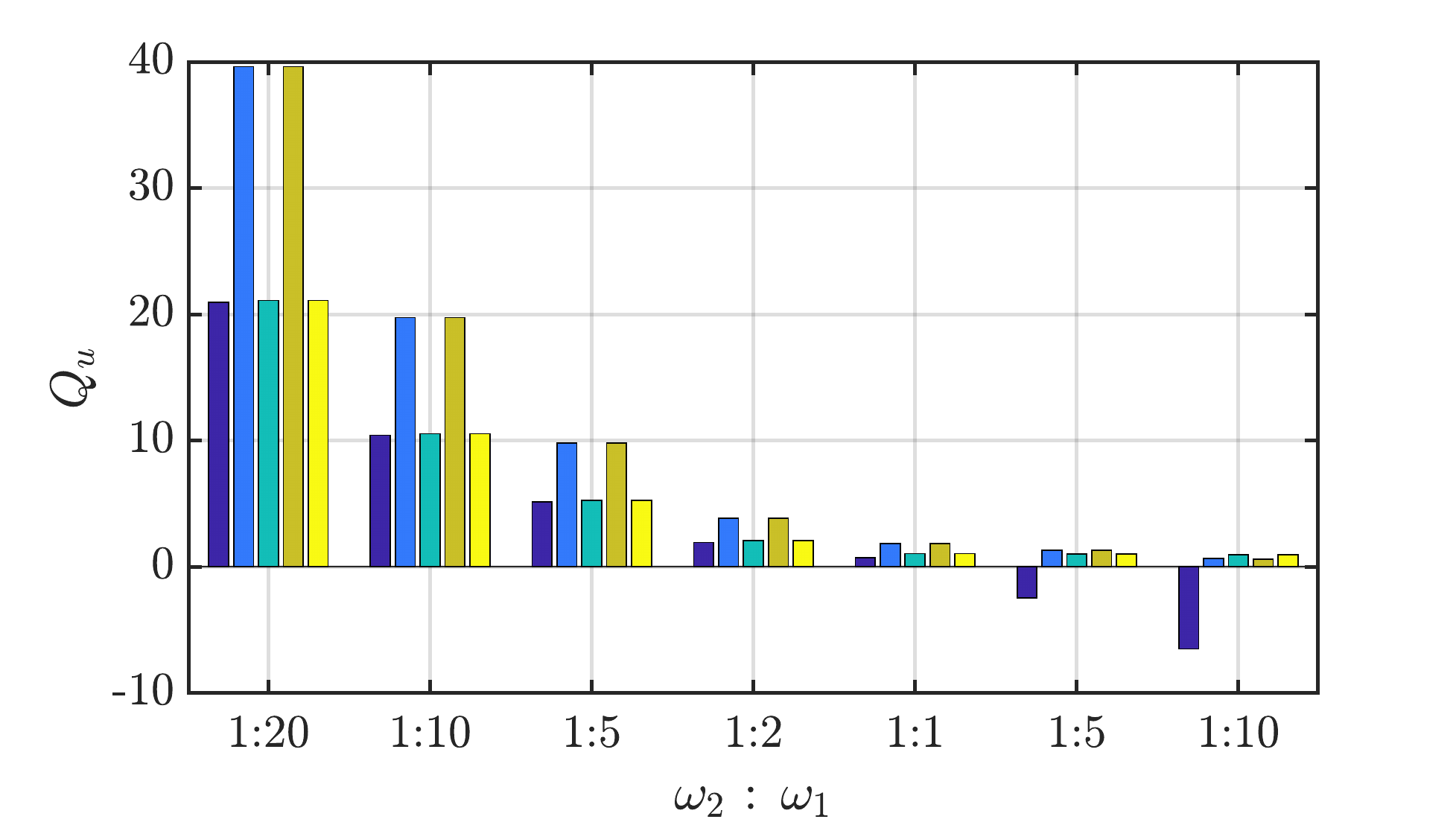}\label{fig:roundquw}}
\subfigure[]
{\includegraphics[width=0.32\textwidth]{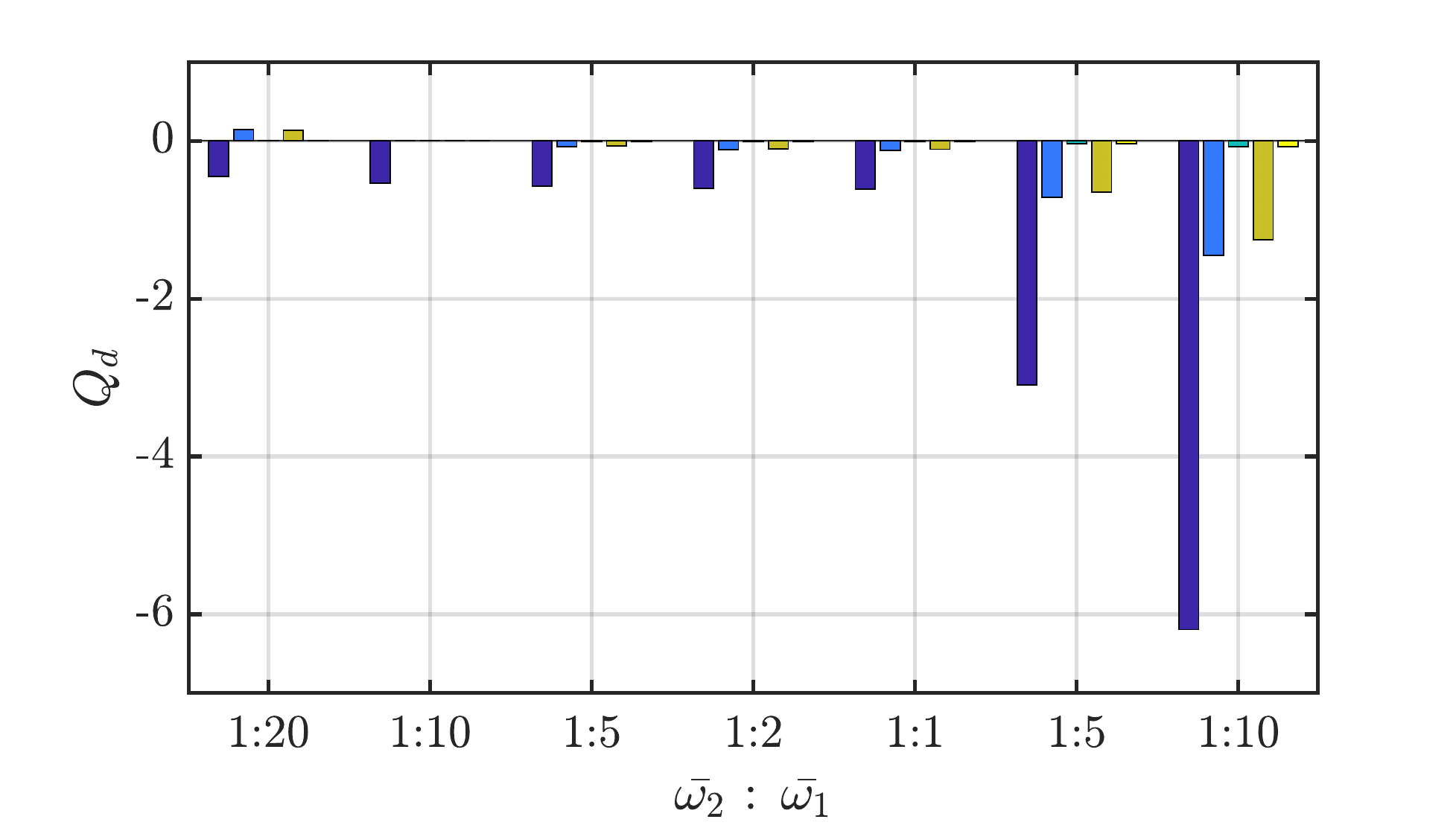}\label{fig:roundqdw}}

\caption{\journal{Performance comparison between \textit{FAIR} and baseline in the roundabout scenario. (a) Road topology and example vehicle trajectories at the roundabout; (b) Comparison of channel utilization with variant number of UCVs and DCVs; (c) and (d) Comparison of average frame rate with variant number of UCVs and DCVs; (e) and (f) Comparison of optimality with variant number of UCVs and DCVs; (g) and (h) Comparison of optimality with variant user preference.}}
\label{fig:round}   
\end{figure*}

\begin{figure*}[t]
\centering
\subfigure[]
{\includegraphics[width=0.5\textwidth]{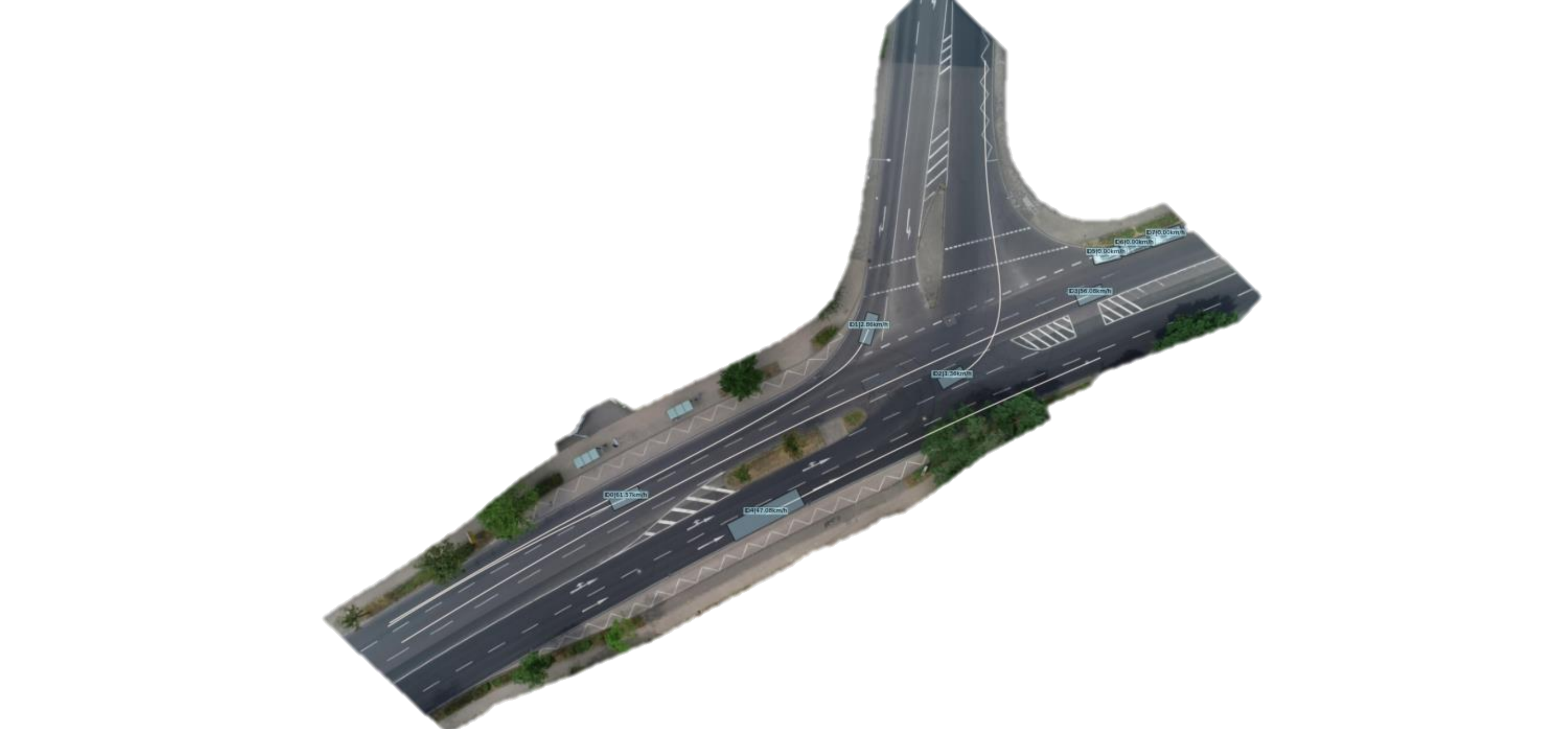}\label{fig:Indmap}}
\subfigure[]
{\includegraphics[width=0.32\textwidth]{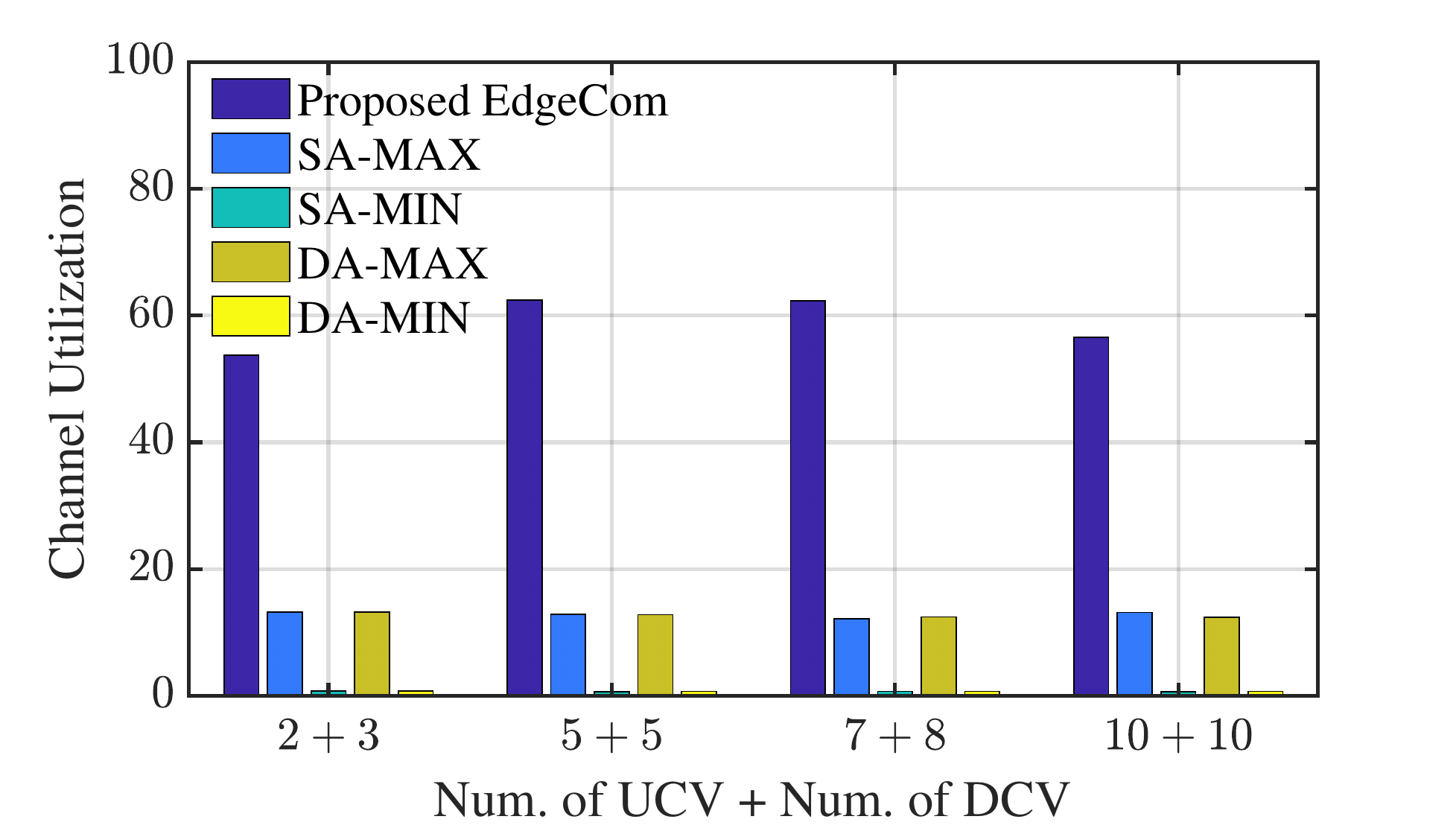}\label{fig:IndUti}}
\subfigure[]
{\includegraphics[width=0.32\textwidth]{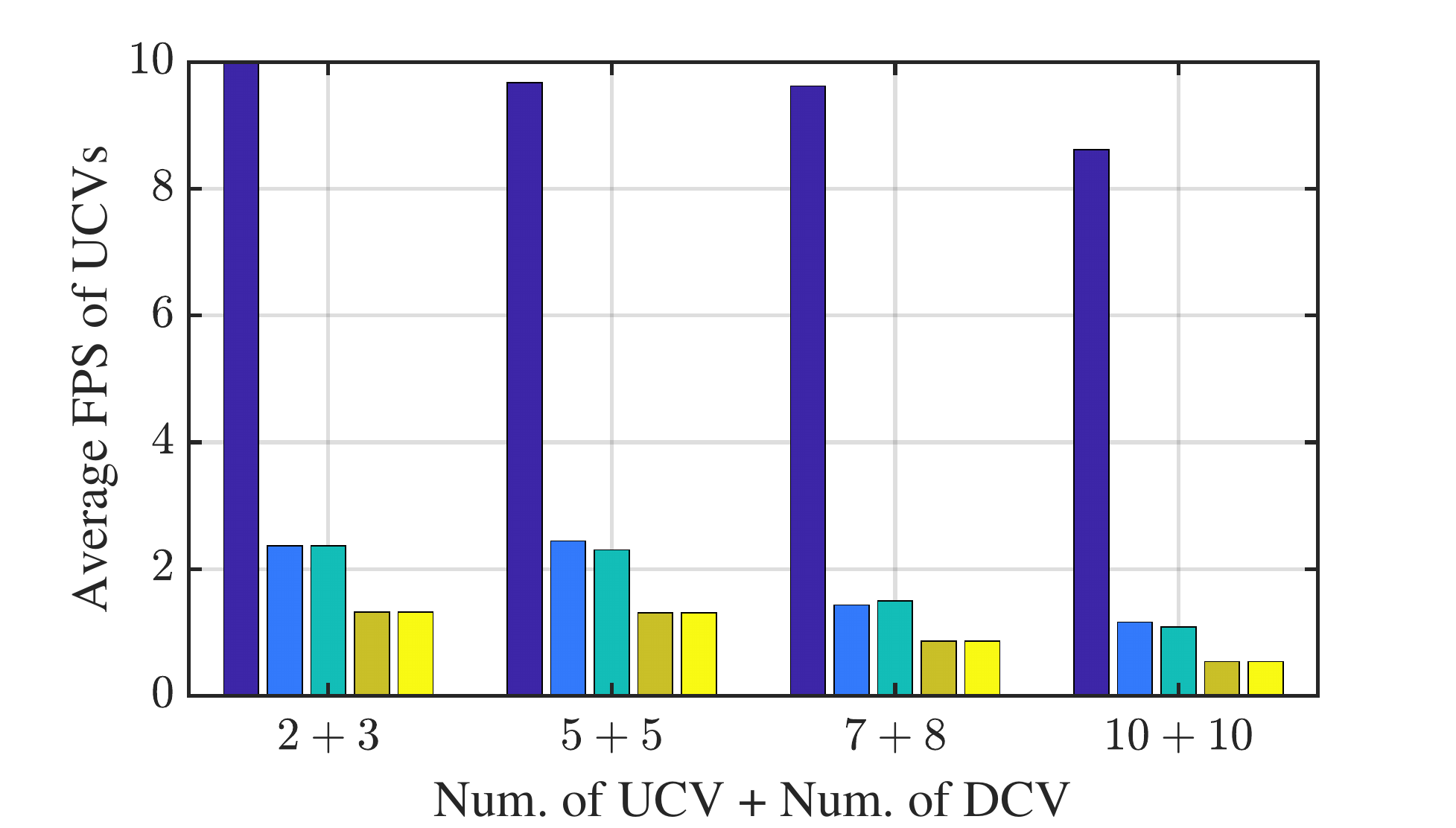}\label{fig:Indfpsu}}
\subfigure[]
{\includegraphics[width=0.32\textwidth]{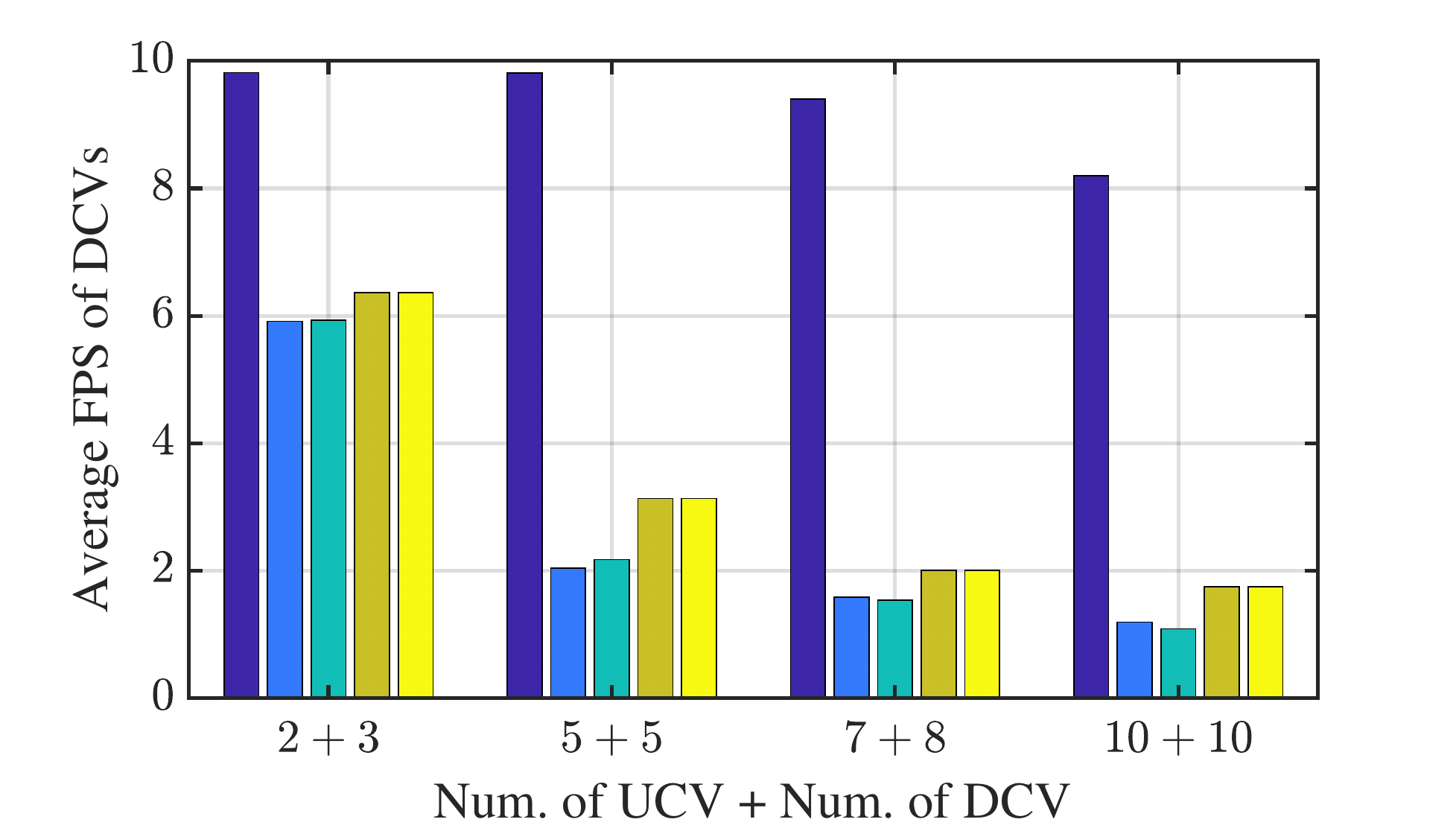}\label{fig:Indfpsd}}
\subfigure[]
{\includegraphics[width=0.32\textwidth]{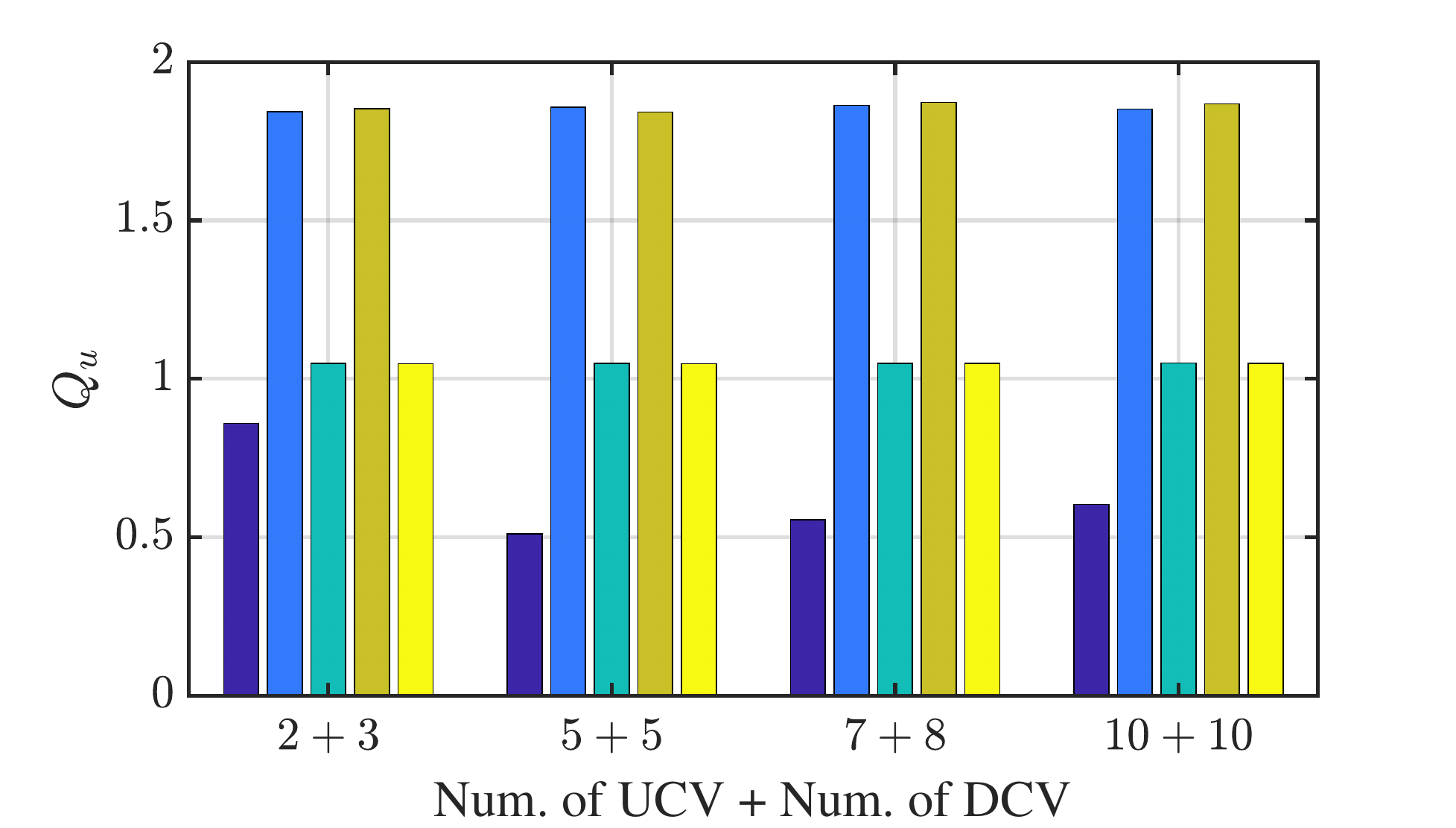}\label{fig:Indqunum}}
\subfigure[]
{\includegraphics[width=0.32\textwidth]{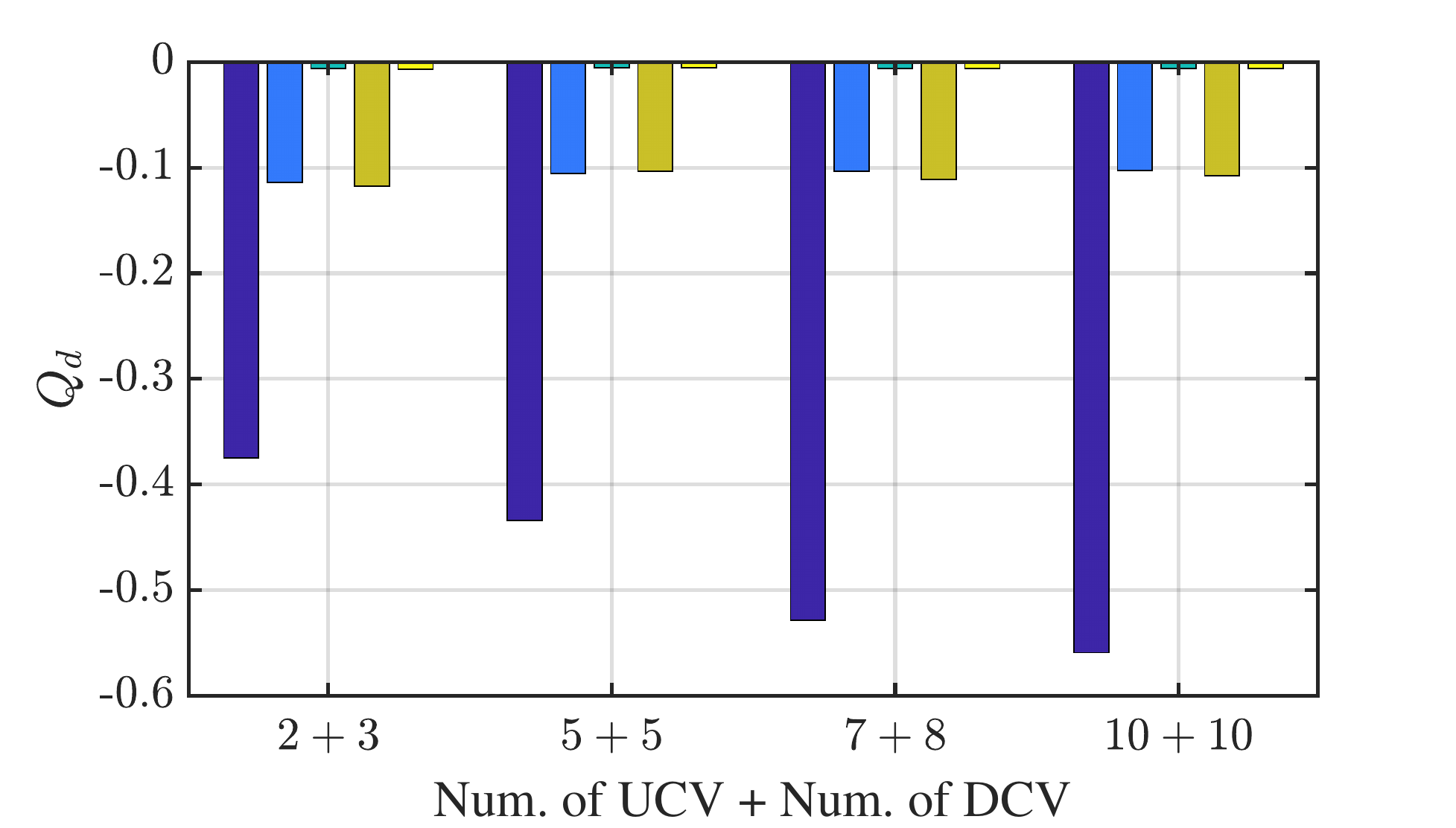}\label{fig:Indqdnum}}
\subfigure[]
{\includegraphics[width=0.32\textwidth]{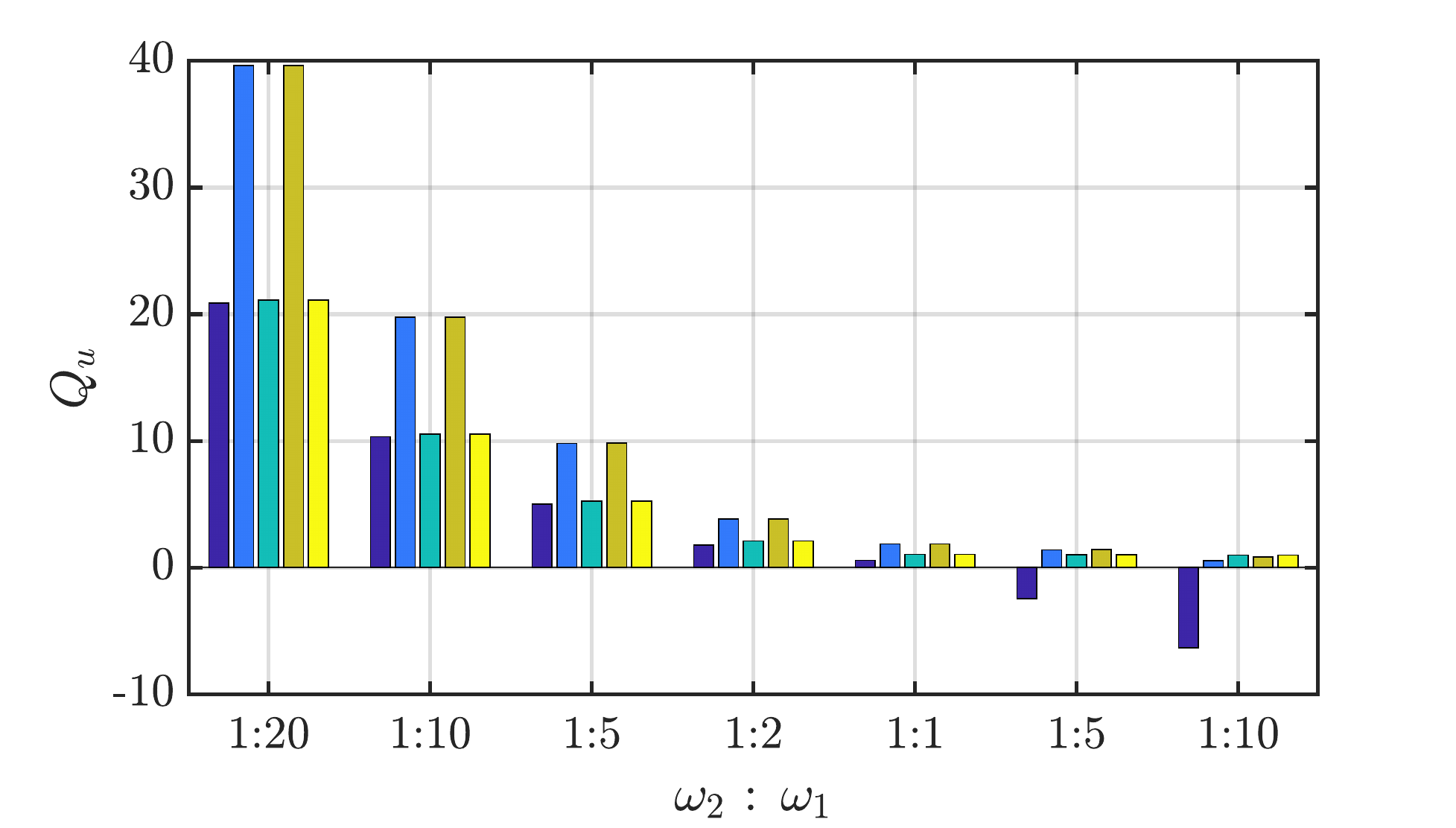}\label{fig:Indquw}}
\subfigure[]
{\includegraphics[width=0.32\textwidth]{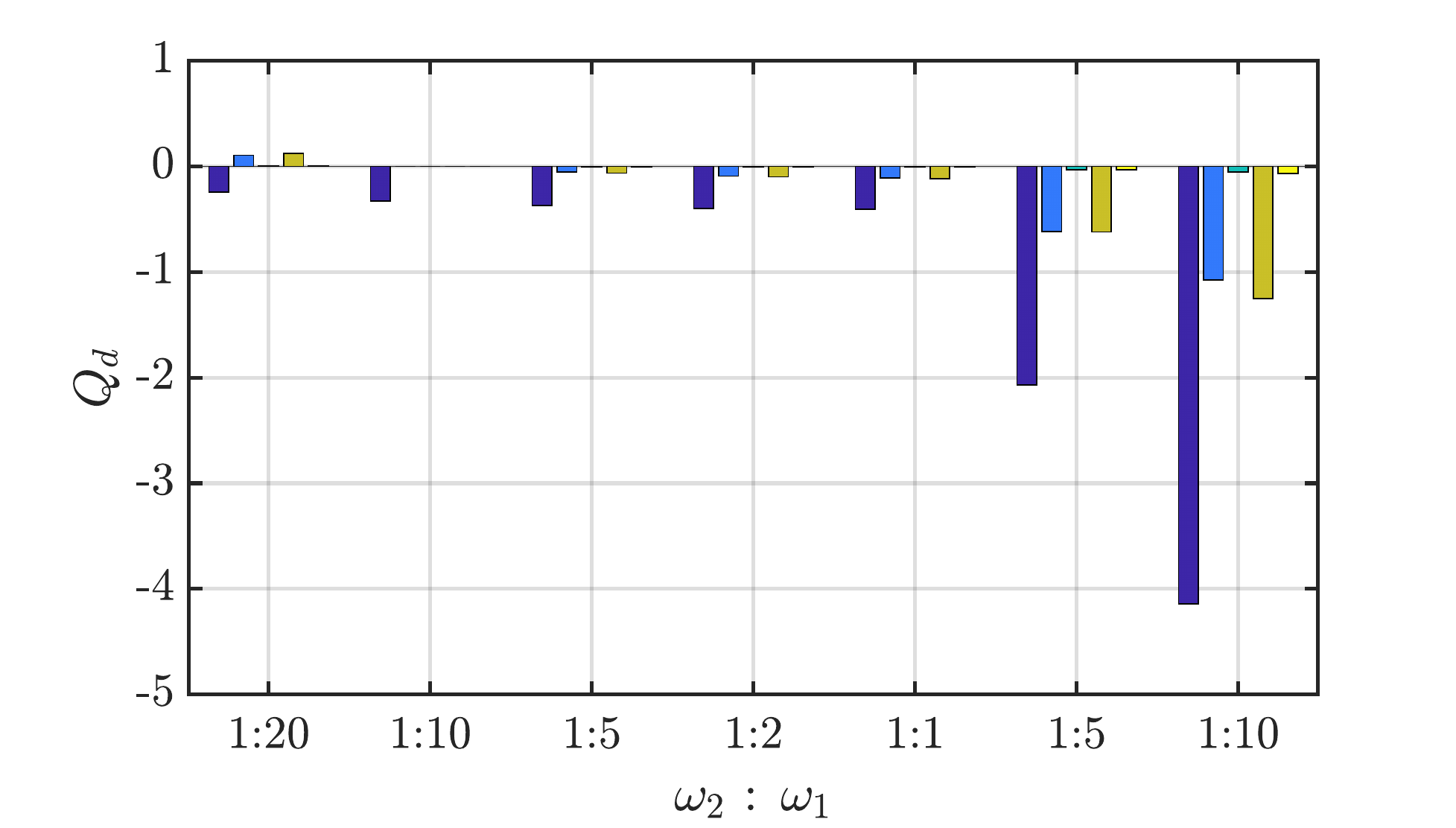}\label{fig:Indqdw}}

\caption{\journal{Performance comparison between \textit{FAIR} and baseline in the suburb intersection scenario. (a) Road topology and example vehicle trajectories at the intersection; (b) Comparison of channel utilization with variant number of UCVs and DCVs; (c) and (d) Comparison of average frame rate with variant number of UCVs and DCVs; (e) and (f) Comparison of optimality with variant number of UCVs and DCVs; (g) and (h) Comparison of optimality with variant user preference.}}
\label{fig:ind}   
\end{figure*}

\journal{This section presents the performance evaluation of our proposed \textit{FAIR} scheme through vehicle trajectory datasets collected in multiple real-world traffic scenarios and road topology. We first introduce the real-world dataset we used in the simulation. Then, we evaluate the average FPS, per frame energy consumption, channel utilization, and average obtained frame resolution of the proposed \textit{FAIR} under variant configurations through data-driven simulations.

\subsection{Trajectory Dataset}
To validate our proposed \textit{FAIR} can achieve network performance improvement for connected vehicles, we conduct extensive end-to-end simulations with real-world vehicle traces collected by drones \cite{rounDdataset, inDdataset}. The datasets include multiple traffic scenes and road topology, e.g., roundabout, intersection, and straight road. The datasets consist of trajectories of over 20,000 road users, including cars, trucks, vans, buses, pedestrians, bicycles and motorcycles, where the trajectories are generated from the videos recorded by drones in $25$Hz at different locations. Since our work is designed for connected vehicles, we filter out pedestrians, bicycles, and motorcycles in the datasets.

\subsection{Performance Evaluation of FAIR}
We conduct simulations with one edge server and a number of connected vehicles that have either downlink services (i.e., downloading entertainment videos) or uplink services (i.e., intelligent driving), or both. As we present in Section \ref{sc:Proposed Scheme}, a vehicle can obtain uplink and downlink traffic simultaneously. Furthermore, we utilize a path-loss model to simulate the network dynamics caused by the mobility of vehicles:
\begin{equation}
\label{eq:path-loss}
P_L = 20\log_{10}{f} + 10n\log_{10}{d} - 24 (dB),
\end{equation}
where $d$ is the distance between the edge server and a vehicle (UCV or DCV); $P_L$ describes the radio frequency (RF) propagation path-loss, which is determined by $d$, $f$ (i.e., carrier frequency), and $n$ (i.e., path-loss exponent). $n$ and $f$ are set to $6$ and $2400$ MHz in our simulation environment. In addition, the Signal-to-Noise Ratio (SNR) is introduced and calculated to estimate the real-time data transmission rate of each vehicle-to-server connection. The correlation between SNR and data rate is shown in Table \ref{tb:tb2}. $\theta = 50^\circ$, $l=55$ m \cite{BOSCH}, $fps_{0}=10$ Hz \cite{lin2018architectural,geiger2012we} are the configurations of the camera sensors equipped on each connected vehicle. The support image frame resolutions for offloading are $k\cdot s = \{640\cdot 480, 480\cdot 480, 320 \cdot 480, 320 \cdot 320, 224 \cdot 320, 224\cdot 224, 128 \cdot 224, 128 \cdot 128\}$ pixels. The support video frame resolution for downloading are $\Bar{k} \cdot \Bar{s} = \{640\cdot 480, 480\cdot 480, 320 \cdot 480, 320 \cdot 320, 224 \cdot 320, 224\cdot 224, 128 \cdot 224, 128 \cdot 128\}$ pixels. \revision{The number of bits carried by one pixel is set as $8$.} The default user preference are $\omega_{1} = 1, \omega_{2} = 1$, and $\Bar{\omega_{1}} = 1, \Bar{\omega_{2}} = 1$. \revision{The edge server is set to update the DOP reservation once every $T=100$ ms. The speed threshold $V_{highway}$ in case 1 is set as $26.8$ m/s. The trigger $\beta$ of case 2, continous unallocated, is set as $3$. UCVs classified as case 3 will be allowed to offload every $10\cdot T$.}

\begin{table}[t]
\centering
\caption{Data rate table of 802.11n ($4$ spatial streams)}
\begin{tabular}{|c|c|c|c|c|c|c|c|c|}
\hline
Min. SNR (dBm) & 2  & 5 & 9   & 11 & 15 & 18 & 20 & 25 \\\hline
Data rate (Mbps)& 29 & 58 & 87 & 116 & 173  & 231  & 260 &  289 \\\hline
\end{tabular} 
\label{tb:tb2}
\end{table}

\begin{table}[t]
\centering
\caption{Power Consumption and Duration of Promotion \& Tail Phases in Data Transmission.}
\begin{tabular}{|c|c|c|c|}
\hline
$P_{pro}$ (W) & $t_{pro}$ (s) & $P_{tail}$ (W) & $t_{tail}$ (s)\\ \hline
$1.97\pm 0.08$ & $0.034\pm 0.004$ & $1.61\pm 0.15$ & $0.21\pm 0.02$\\ \hline
\end{tabular}
\label{tb:promo&tail}
\end{table}

We compare our propose \textit{FAIR} with the following algorithms:
\begin{itemize}
    \item \textit{SA $+$ MAX:} The edge server and connected vehicles obtain the \underline{s}ame \underline{A}IFS. Connected vehicles adopt the highest frame resolution for their connected services to achieve \underline{max}imum accuracy.
    \item \textit{SA $+$ MIN:} The vehicle adopts the lowest frame resolution for its connected services to achieve the \underline{min}imum latency.
    \item \textit{DA $+$ MAX:} The edge server obtains a shorter AIFS than connected vehicles, i.e., \underline{d}ifferent \underline{A}IFS. The vehicle adopts the highest frame resolution for its connected services.
    \item \textit{DA $+$ MIN:} The vehicle adopts the lowest frame resolution for its connected services.
\end{itemize}

\textbf{From system perspective.}
We evaluate the performance of \textit{FAIR} at both roundabout and intersection. We aim to study: 1) what's the performance of \textit{FAIR} compared to other solutions; 2) how does \textit{FAIR} perform in different traffic scenes; 3) whether \textit{FAIR} can effectively scale under the different number of connected vehicles. Figs. \ref{fig:roundmap} and \ref{fig:Indmap} illustrate the road topology of roundabout and intersection adopted in our simulations. In particular, the roundabout in Fig. \ref{fig:roundmap} is four-armed and connects a highway with Aachen in Germany, where it obtains the highest traffic volume among all the recorded scenes in the datasets \cite{rounDdataset}. The average traffic speed is $7.41 m/s$, and the speed of vehicles is in the range of $[0, 17.72] m/s$. The intersection in Fig. \ref{fig:Indmap} is located in suburban area of Aachen and is a T-junction, where the major arterial is straight and has the right of way, and there is a left turn lane into the side road\cite{inDdataset}. The average traffic speed is $12.32 m/s$, and the speed of vehicles is in the range of $[0, 25.13] m/s$.

Figs. \ref{fig:roundUti} and \ref{fig:IndUti} show the channel utilization performance of different algorithms under different number of UCVs and DCVs. We can see that our proposed \textit{FAIR} outperforms other algorithms and always achieves the best channel utilization performance, which demonstrates our analysis in Section \ref{ssc:adaptation}. For example, in the roundabout scenario, \textit{FAIR} improves $4.6$ times channel utilization in average compared to SA-MAX and DA-MAX, which validates \textit{FAIR} can effectively utilize the network resource and enhance the resource accessibility through proactive scheduling and avoiding reactive channel contention among vehicles.

Figs. \ref{fig:roundfpsu}, \ref{fig:roundfpsd} and Figs. \ref{fig:Indfpsu}, \ref{fig:Indfpsd} evaluate the average frame rate achieved by UCVs and DCVs at roundabout and intersection, respectively. We illustrate how \textit{FAIR} performs with different number of UCVs and DCVs. We observe that \textit{FAIR} always assures both UCVs and DCVs of obtaining a high frame rate (i.e., close to $10$ Hz). In addition, compared to other algorithms, \textit{FAIR} significantly increases the frame rate of both UCVs and DCVs. For instance, in the intersection scenario, it increases the average frame rate of UCVs by $6.7$ and $11.2$ times, and the average frame rate of DCVs by $5.9$ and $4.7$ times, compared to SA and DA, respectively, when the number of UCVs and DCVs is $7 + 8$. In the roundabout scenario, \textit{FAIR} increases the average frame rate of UCVs by $13.7$ and $30.2$ times, and the average frame rate of DCVs by $12.6$ and $8.3$ times, compared to SA and DA, respectively, when the number of UCVs and DCVs is $10 + 10$. The larger number of UCVs and DCVs, the higher increment of frame rate \textit{FAIR} can achieve. 
These results validate that \textit{FAIR} provides an impartial network resource allocation for uplink and downlink traffic, and is scalable under the different number of connected vehicles and workload.

Figs. \ref{fig:roundqunum}, \ref{fig:roundqdnum} and Figs. \ref{fig:Indqunum}, \ref{fig:Indqdnum} further validate the optimality of our proposed \textit{FAIR} with the variant number of UCVs and DCVs at roundabout and intersection, respectively. While Figs. \ref{fig:roundquw}, \ref{fig:roundqdw} and Figs. \ref{fig:Indquw}, \ref{fig:Indqdw} validate the optimality of \textit{FAIR} with the variant user preference (i.e, $\omega_{1}, \omega_{2}$) at roundabout and intersection, respectively. We observe that \textit{FAIR} always obtains the minimal $Q_{u}$ and $Q_{d}$ compared to other algorithms at both roundabout and intersection.

\begin{figure*}[t]
\centering
\subfigure[Vehicle speed]
{\includegraphics[width=0.8\textwidth]{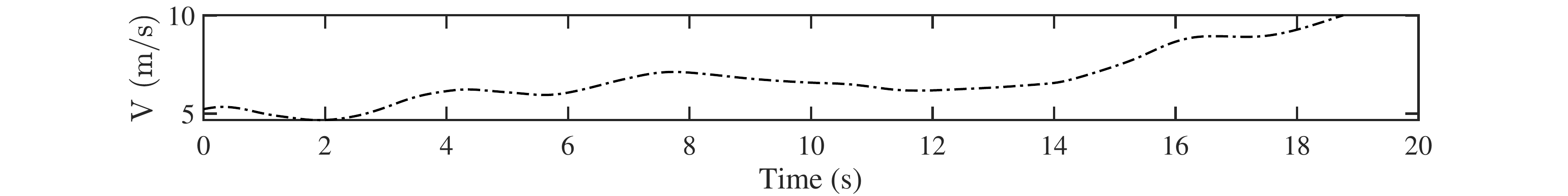}\label{fig:speed}}
\subfigure[Signal-to-noise ratio]
{\includegraphics[width=0.8\textwidth]{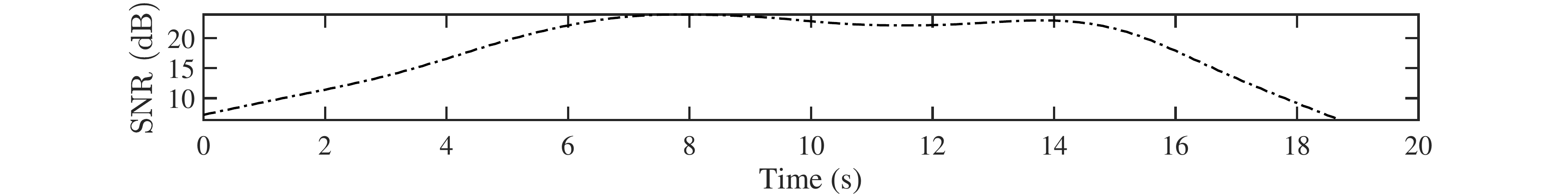}\label{fig:snr}}
\subfigure[Data rate]
{\includegraphics[width=0.8\textwidth]{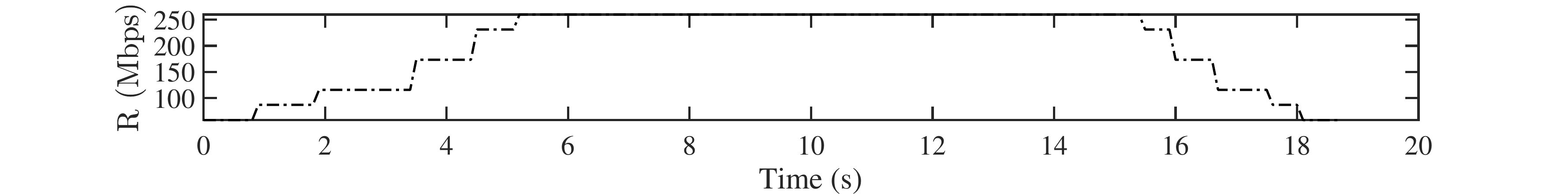}\label{fig:datarate}}
\subfigure[Allocated DOP]
{\includegraphics[width=0.8\textwidth]{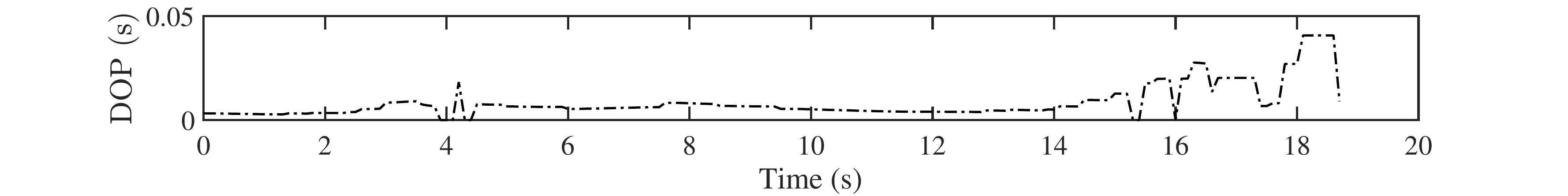}\label{fig:dop}}
\subfigure[Per frame energy consumption]
{\includegraphics[width=0.8\textwidth]{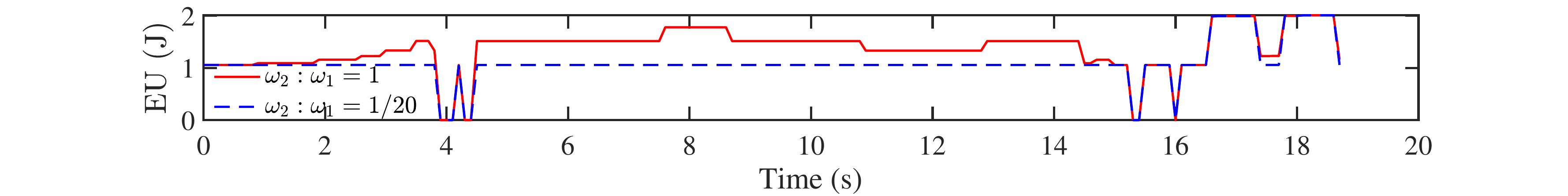}\label{fig:energy}}
\subfigure[Utilization of allocated DOP]
{\includegraphics[width=0.8\textwidth]{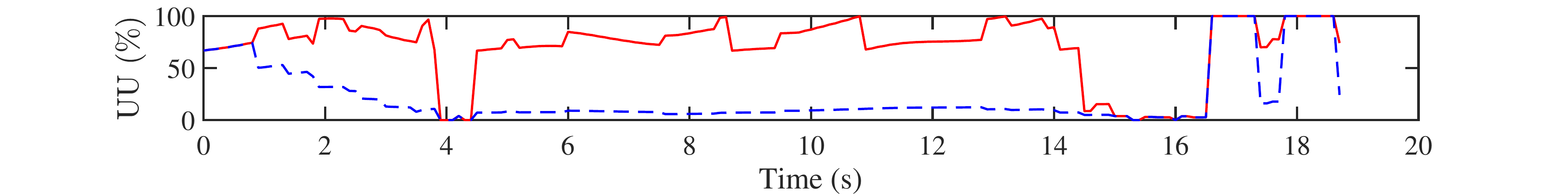}\label{fig:uu}}
\subfigure[Frame resolution]
{\includegraphics[width=0.8\textwidth]{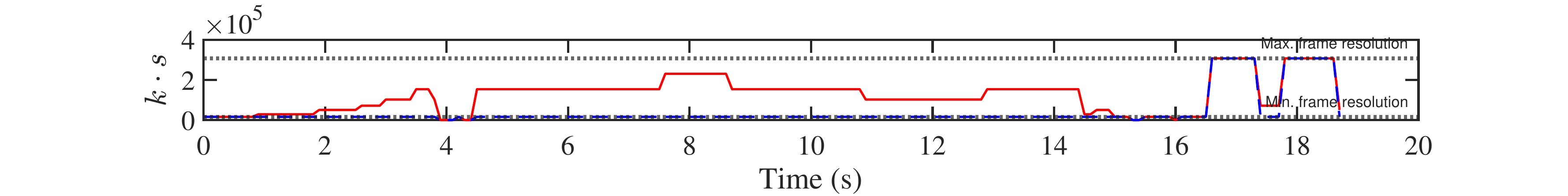}\label{fig:resolution}}
\caption{\journal{The performance results of a connected vehicle. (a)~(d) show the selected connected vehicle's instantaneous speed, signal-to-noise ratio, obtained data rate, and allocated DOP over the time; (e)~(g) illustrate the comparison of two user preference settings (e.g., $\omega_{2}:\omega_{1} = 1$ and $\omega_{2}:\omega_{1} = 20$) in terms of the per frame energy consumption, utilization of the allocated DOP, and offloaded image frame resolution over the time.}}
\label{fig:UCV}   
\end{figure*}

\textbf{From individual connected vehicle perspective.} In Fig. \ref{fig:UCV}, we illustrate how \textit{FAIR} makes the DOP allocation and connected service offloading intelligent under the environment dynamics and varying user preference. We randomly select a UCV in the simulation environment, where the road topology is the roundabout and the number of UCVs and DCVs is $10$ and $10$, respectively. Fig. \ref{fig:speed} illustrates the instantaneous speed of the UCV. Fig. \ref{fig:snr} shows the signal-to-noise (SNR) ratio on the side of the UCV, which also indicates the UCV is in the range of the edge about $19$ s. Fig. \ref{fig:datarate} shows the data rate of the UCV and it also demonstrates the correlation between SNR and data rate. Fig. \ref{fig:dop} illustrates the allocated DOP for the UCV over the time. We can observe that the allocated DOP increases when the UCV's speed is higher (the allocated DOP is also highly affected by other connected vehicles in the environment, which cannot be demonstrated in this figure).

Figs. \ref{fig:energy}, \ref{fig:uu}, and \ref{fig:resolution} show the real-time performance of the UCV in terms of per frame energy consumption, $EU$ (i.e., defined in \ref{eq:EU}), utilization of the allocated DOP, $UU$ (i.e., defined in \ref{eq:Q1}), and the optimal frame resolution determined by Algorithm \ref{alg:2}, \revision{with two different user preference settings $\omega_{2}:\omega_{1} = 1:1$ or $1:20$. The red solid line depicts the real-time performance of the UCV with $\omega_{2}:\omega_{1} = 1:1$; while the blue dashed line depicts the real-time performance of the UCV with $\omega_{2}:\omega_{1} = 1:20$.} The energy consumption of transmitting an image frame, $EU_{tr}$, is calculated based on Table \ref{tb:promo&tail} and the model $P_{tr} = 0.01821\cdot R + 0.7368$ \cite{wang2020user, wang2022leaf, wang2019energy}. The energy consumption of sampling an image frame, $EU_{cam}$, is calculated by $EU_{cam} = -1.772\times 10^{-17}\cdot (k\cdot s)^3 + 7.491\times 10^{-12}\cdot (k\cdot s)^2 + 2.379\times 10^{-6}\cdot (k\cdot s) + 0.6068$ \cite{wang2020user, wang2022leaf, wang2019energy}. We can see that if the user prefers a lower energy consumption (i.e., a larger $\omega_{1}$), \textit{FAIR} will be prone to select a lower frame resolution for sampling and offloading, as shown in Fig. \ref{fig:resolution}. By lowering the frame resolution, the UCV can save up to $40.6\%$ energy consumption per frame. While if the user prefers a higher frame resolution ((i.e., a larger $\omega_{2}$)), the utilization of the allocated DOP will raise significantly, as depicted in Fig. \ref{fig:uu}. These results indicate that our proposed \textit{FAIR} can intelligently optimize the frame resolution selection based on the user preference.
}

\section{Conclusion}
\label{sc:Conclusion}
In this paper, we proposed \textit{FAIR}, an end-to-end automotive edge networking system, that can provide fast, scalable, and impartial connected services for intelligent vehicles with edge computing. This is the first work, to the best of our knowledge, that systematically address the issue of the asymmetrical resource allocation for uplink and downlink connects in traditional wireless networks. We designed a symmetrical resource allocation algorithm to proactively and periodically reserve dedicated service resource for each intelligent vehicle under environment dynamics. A service adaptation algorithm was designed to dynamically adjust configurations of environmental sensing and frame resolution according to the reserved service period and user preference. The extensive simulation results demonstrated \textit{FAIR} can significantly improve the fairness, scalability, and reliability compared to existing solutions in a variety of traffic scenes.

\bibliographystyle{IEEEtran}
\bibliography{reference}
\end{document}